\documentclass[12pt]{article}
\usepackage{graphicx}
\usepackage{amsmath}
\usepackage{amsfonts}
\usepackage{amssymb}

\begin{document}

\title{Extended Thermodynamics for dense gases and macromolecular fluids, obtained through a non relativistic limit}
\author{M.C. Carrisi, M.A. Mele, S. Pennisi}
\date{}
\maketitle \vspace{0.5 cm}
 \small {\em \noindent Dipartimento di Matematica ed Informatica,
Universit\`{a} di Cagliari, Via Ospedale 72,\,\ 09124 Cagliari, Italy; \, e-mail:
cristina.carrisi@tiscali.it; spennisi@unica.it} \\

\begin{abstract}
In this paper we consider the 14 moments model of Extended Thermodynamics for dense gases
and macromolecular fluids. Solutions of the restrictions imposed by the entropy principle
and that of Galilean relativity for such a model have been until now obtained in
literature only in an approximate manner up to a certain order with respect to
thermodynamic equilibrium; for more restrictive models they have been obtained up to
whatever order, but by using  Taylor expansions around equilibrium and without proving
convergence. Here we have found an exact solution without using expansions. The idea has
been to write firstly a relativistic model, for which it is easy to impose the
Einsteinian relativity principle, and then taking its non relativistic limit.
\end{abstract}
\section{Introduction}
The 14 moments model of Extended Thermodynamics for dense gases and macromolecular fluids
was firstly studied by Kremer in \cite{1}, up to second order with  respect to
equilibrium. The balance equations to describe this model are
\begin{eqnarray}\label{0.1}
& {}& \partial_t F + \partial_k F_k = 0 \, ,   \quad \quad \quad \,
\partial_t F_i + \partial_k G_{ki} = 0 \, , \quad
\partial_t F_{ij} + \partial_k G_{kij} = P_{<ij>} \, , \\
& {}& \nonumber
\partial_t F_{ill} + \partial_k G_{kill} =   P_{ill} \, , \quad
\partial_t F_{iill} + \partial_k G_{kiill} =   P_{iill} \, ,
\end{eqnarray}
where the independent variables are $F$, $F_{i}$, $F_{i j}$, $F_{ill}$, $F_{iill}$ and
are symmetric tensors. $P_{< i j>}$, $P_{ ill }$, $P_{iill}$ are productions and they too
are symmetric tensors.The fluxes  $G_{ki}$, $G_{kij}$, $G_{kill}$, $G_{kiill}$ are
constitutive functions and are symmetric over all indexes, except for $k$. \\
In ideal gases, we have also the conditions $G_{ki}=F_{ki}$, $G_{ill}=F_{ill}$,
$G_{iill}=F_{iill}$ and, moreover, $G_{kij}$ and $G_{kill}$ are symmetric over all
couples of indexes; then the present case is less restrictive. \\
We want that our system $(\ref{0.1})$ is a symmetric hyperbolic one, with all consequent
nice mathematical properties. To this end, we impose the entropy principle; in the next
section it will be proved that it is equivalent to the following equations
\begin{eqnarray}\label{mel1}\nonumber
F&=&\frac{\partial h'}{\partial\lambda},\quad F_{i}=\frac{\partial
h'}{\partial\lambda_{i}},\quad F_{il}=\frac{\partial
h'}{\partial\lambda_{il}},\,\\\nonumber\quad F_{ill}&=&\frac{\partial
h'}{\partial\lambda_{ill}},\quad F_{iill}=\frac{\partial h'}{\partial\lambda_{iill}} \,
,\\~
\\
G_{k}&=&\frac{\partial\phi'_{k}}{\partial\lambda},\quad
G_{ki}=\frac{\partial\phi'_{k}}{\partial\lambda_{i}},\quad
G_{kil}=\frac{\partial\phi'_k}{\partial \lambda_{il}},\,\nonumber
\\\nonumber
G_{kill}&=&\frac{\partial\phi'_{k}}{\partial\lambda_{ill}},\quad
G_{kiill}=\frac{\partial\phi'_{k}}{\partial\lambda_{iill}} \, ,
\end{eqnarray}
where we have taken into account the definition $G_{k}=F_{k}$,  from which the following
compatibility conditions holds
\begin{eqnarray}\label{mel2}
\frac{\partial\phi'_{k}}{\partial\lambda}&=&\frac{\partial h'}{\partial\lambda_{k}} \, .
\end{eqnarray}
Moreover, also in the next section, it will be proved that the Galilean relativity
principle is equivalent to the following two other conditions
\begin{eqnarray}\label{mel3}
0&=& \frac{\partial h'}{\partial\lambda} \lambda_{i}+2\lambda_{ij}\frac{\partial
h'}{\partial\lambda_{j}}+ \lambda_{jpp}\left(\frac{\partial h'}{\partial\lambda_{rs}}
\delta_{rs} \delta_{ij}+2\frac{\partial
h'}{\partial\lambda_{ij}}\right)+ 4\lambda_{ppqq}\frac{\partial h'}{\partial\lambda_{ill}} \,\\
0&=& \frac{\partial \phi'_{k}}{\partial\lambda} \lambda_{i}+2\lambda_{ij}\frac{\partial
\phi'_{k}}{\partial\lambda_{j}}+ \lambda_{jpp}\left(\frac{\partial
\phi'_{k}}{\partial\lambda_{rs}} \delta_{rs} \delta_{ij}+2\frac{\partial
\phi'_{k}}{\partial\lambda_{ij}}\right)+ 4\lambda_{ppqq}\frac{\partial
\phi'_{k}}{\partial\lambda_{ill}} +{h'}\delta_{ik} \, . \nonumber
\end{eqnarray}
In other words, we have to found  $h'$ and $\phi'_{k}$ satisfying the equations
$(\ref{mel2})$ and $(\ref{mel3})$; after that, $(\ref{mel1})_{1-5}$ are useful to obtain
$\lambda$, $\lambda_{i}$, $\lambda_{ij}$, $\lambda_{ill}$, $\lambda_{iill}$ as functions
of our independent variables $F$, $F_{i}$, $F_{i j}$, $F_{ill}$, $F_{iill}$. Lastly,
$(\ref{mel1})_{7-10}$ will give the constitutive functions $G_{ki}$,
$G_{kij}$, $G_{kill}$, $G_{kiill}$. \\
Now, the restrictions $(\ref{mel2})$ and $(\ref{mel3})$ have been imposed by Kremer in
\cite{1} only up to secondo order with respect to equilibrium. In \cite{2} Carrisi and
Pennisi have imposed them up to whatever order (for the more restrictive model of ideal
gases), but by using Taylor' s expansions around  equilibrium, without worrying about
convergence problems. Here we have found the exact solution without using expansions. The
idea to obtain this result has been the following
\begin{itemize}
  \item Firstly we have assumed a relativistic model for which it is easy to impose the
  Einsteinian relativity principle; the results for this model can be found in section 3;
  \item After that, we have shown, in sect. 4, how to take the non relativistic limit of
  the model in sect.3; to this end we have used a methodology which is easy to find for
  ideal gases because  in this case we have suggestions from the kinetic theory of gases; we have
  adopted this methodology also for our more general case. Obvious, the validity of this
  assumption has to be tested at the end by verifying that our results satisfy truly the
  eqs. $(\ref{mel2})$ and  $(\ref{mel3})$. We will be obtain that eqs. $(\ref{mel3})$ are
  identically satisfied. Instead of this, the condition  $(\ref{mel2})$ will have still to be
  imposed; it is the only condition for which there is no correspondence between the
  relativistic case and the classical one.
  \item Then, in sect. 5 we have taken effectively this limit and found that
  $h'$ and $\phi'_{k}$ are determined by the following eqs.
  $(\ref{ange2})$ and $(\ref{ange2biss})$, except for 4 scalar arbitrary functions $H_0$, $H_1$, $H_2$, $H_3$ which depend on
  the scalars $(\ref{ange3})$.
\end{itemize}
There remain the further condition $(\ref{mel2})$, the requirement of convexity for the
function $h'$ and the problem of subsystems. We have exploited them, but don' report here
the results, for the sake of brevity. We assure only that we have found the exact
solution of $(\ref{mel2})$ without using expansions. \\
We close this section by reporting the results of sect. 5; in this way it will be not
necessary to search them throughout the paper and they will be available for the
applications. They are
\begin{eqnarray}\label{ange2}
\phi'^{k}&=&H_{0}V^k_0+H_{1} V^k_1 + H_{2}V^k_2 + H_{3} V^k_3 \, , \\
h'&=&8H_{0}X_1 - H_{1} X_2- \frac{2}{3}H_2 X_3- \frac{1}{2} H_{3} X_4 \, , \nonumber
\end{eqnarray}
with
 \begin{eqnarray}\label{ange2biss}
V^k_0&=&-2\lambda_{kll} \\
V^k_1&=&-2\lambda_{kh}\lambda_{hll}+4\lambda_{ppll}\lambda_{k}+\frac{4}{5}\lambda_{ll}\lambda_{kll} \,\nonumber \\\
V^k_2&=&-2\lambda^{2}_{kh}\lambda_{hll}+\frac{6}{5}\lambda_{ll}\lambda_{ka}\lambda_{all}+4\lambda_{ka}\lambda_{a}\lambda_{ppll}+\,\nonumber \\\
&-&\frac{11}{25}\lambda^{2}_{ll}\lambda_{kll}-\lambda_{kll}\lambda_{a}\lambda_{all}+\lambda_{k}\lambda_{all}\lambda_{all}+\,\nonumber \\\
&+& (tr
\lambda^{2}_{ab})\lambda_{kll}-\frac{12}{5}\lambda_{ppll}\lambda_{ll}\lambda_{k} \nonumber \\
V^k_3&=& 2\lambda_{ppll} \left( 2\lambda^{2}_{kh}\lambda_{h} - tr \lambda^{2}_{ab}
\lambda_{k}-\frac{8}{5}\lambda_{ll}\lambda_{ka}\lambda_{a}+
\frac{17}{25}\lambda^{2}_{ll}\lambda_{k} \right) +  \nonumber \\
&+&(\lambda_{kh}\lambda_{h})(\lambda_{all}\lambda_{all}) - \frac{4}{5}\lambda_{ll}
(\lambda_{all}\lambda_{all})\lambda_{k}-\frac{17}{25}\lambda^{2}_{ll}\lambda_{ka}\lambda_{all}+
\nonumber
\\
&-&(\lambda_{a}\lambda_{all})\lambda_{kb}\lambda_{bll}+(tr
\lambda^{2}_{ab})\lambda_{kc}\lambda_{cll}+ \frac{4}{5}\lambda_{ll}
(\lambda_{a}\lambda_{all})\lambda_{kll}+ \nonumber
\\
&+& \frac{8}{5}\lambda_{ll}\lambda^{2}_{kh}\lambda_{hll}+
\frac{74}{375}\lambda^{3}_{ll}\lambda_{kll}  - \frac{4}{5} \lambda_{ll}(tr
\lambda^{2}_{ab})
\lambda_{kll}+ (\lambda_{ab} \lambda_{all} \lambda_{bll})\lambda_{k} + \nonumber \\
&-&   (\lambda_{ab} \lambda_{a} \lambda_{bll})\lambda_{kll} + \frac{2}{3} (tr
\lambda^{3}_{ab}) \lambda_{kll}-2\lambda^{3}_{kh}\lambda_{hll}  \, , \nonumber
\end{eqnarray}
\begin{eqnarray}\label{ange3}
X_1 &=& \lambda_{ppll} \, , \\
X_2 &=& 2 \lambda_{all}
\lambda_{all}-  \frac{16}{5}\lambda_{ppll}\lambda_{ll} \, ,   \nonumber \\
X_3 &=& 8 \lambda_{ppll} \left( \frac{11}{50} \lambda^{2}_{ll}  - \frac{1}{2} tr
\lambda^{2}_{ab}\right)
+ 2  \lambda_{ab} \lambda_{all} \lambda_{bll} - \frac{6}{5} \lambda_{ll} \lambda_{all} \lambda_{all}  \, ,  \nonumber \\
X_4 &=& 2\lambda^{2}_{ab}\lambda_{all} \lambda_{bll}- tr \lambda^{2}_{ab} \lambda_{cll}
\lambda_{cll} -\frac{8}{5}\lambda_{ll} \lambda_{ab} \lambda_{all} \lambda_{bll}+
\nonumber
\\
& & + \frac{17}{25}\lambda^{2}_{ll}  \lambda_{all} \lambda_{all}+
 8 \lambda_{ppll} \left( - \frac{37}{375}\lambda^{3}_{ll} +
\frac{2}{5} \lambda_{ll}(tr \lambda^{2}_{ab}) - \frac{1}{3} tr
\lambda^{3}_{ab} \right) \, ,  \nonumber \\
X_5 &=& - \frac{2}{5} \lambda^{2}_{ll} +  16 \lambda_{ppll} \Lambda - 4 \lambda_{a}
\lambda_{all} + 2 tr \lambda^{2}_{ab} \, , \nonumber\\
X_6 &=& 4 \Lambda \lambda_{all}
\lambda_{all}+ 8 \lambda_{ppll}
\left( - \frac{4}{5}\Lambda \lambda_{ll} + \frac{1}{2} \lambda_{a}\lambda_{a} \right) +  \nonumber \\
& & + \frac{8}{5} \lambda_{ll}  \lambda_{all}  \lambda_{a} -\frac{4}{5}\lambda_{ll}tr
\lambda^{2}_{ab} + \frac{8}{75}\lambda_{ll}^3 - 4 \lambda_{ab}\lambda_{a} \lambda_{bll}+
\frac{4}{3} tr \lambda^{3}_{ab}  \, , \nonumber
\end{eqnarray}
\begin{eqnarray*}
 X_7 &=& \frac{8}{15}( tr \lambda^{3}_{ab})
\lambda_{ll} - \frac{14}{25} \lambda_{ll}^2 tr \lambda^{2}_{ab} + \frac{46}{375}
\lambda_{ll}^4 +4 \Lambda \lambda_{ab}\lambda_{all}\lambda_{bll} + \nonumber
\\
& &+2 ( tr \lambda^{2}_{ab}) \lambda_{c} \lambda_{cll} - ( \lambda_{a}\lambda_{all})^2
-\frac{12}{5} \Lambda
\lambda_{ll}\lambda_{all}\lambda_{all} +  \nonumber \\
& &+(\lambda_{a}\lambda_{a}) (\lambda_{bll}\lambda_{bll}) - 4
\lambda^{2}_{ab}\lambda_{all}\lambda_{b} + \nonumber \\
& & -8 \lambda_{ppll} \left( \Lambda tr \lambda^{2}_{ab} -\frac{1}{2} \lambda_{ab}
\lambda_{a} \lambda_{b} -\frac{11}{25} \Lambda \lambda_{ll}^2 + \frac{3}{10} \lambda_{ll}
\lambda_{a}
\lambda_{a} \right) + \nonumber \\
& & + \frac{12}{5} \lambda_{ll} \lambda_{ab} \lambda_{a}\lambda_{bll} - \frac{22}{25}
\lambda_{ll}^2 \lambda_{a}\lambda_{all} \, . \nonumber \\
X_8 &=& - \frac{34}{25}\lambda^{2}_{ll} \lambda_{ab}\lambda_{a}\lambda_{bll} + 2 (tr
\lambda^{2}_{ab}) \lambda_{cd}\lambda_{c}\lambda_{dll} + \frac{16}{5} \lambda_{ll}
\lambda_{ab}^2\lambda_{a}\lambda_{bll} +
\\
& &+ \frac{148}{375} \lambda^{3}_{ll} \lambda_{a} \lambda_{all} - \frac{8}{5}\lambda_{ll}
(tr \lambda^{2}_{ab}) \lambda_{c}\lambda_{cll}+ \frac{4}{3} (tr \lambda^{3}_{ab})
\lambda_{c}\lambda_{cll}- 4 \lambda^{3}_{ab} \lambda_{a} \lambda_{bll} +   \\
& & +2 \lambda_{ppll} \left( 2 \lambda^{2}_{ab} \lambda_{a} \lambda_{b}- (tr
\lambda^{2}_{cd}) \lambda_{a} \lambda_{a} - \frac{8}{5} \lambda_{ll} \lambda_{ab}
\lambda_{a}\lambda_{b}+
\frac{17}{25} \lambda_{ll}^2 \lambda_{a} \lambda_{a} \right) + \\
& &+(\lambda_{ab}\lambda_{a}\lambda_{b}) (\lambda_{cll}\lambda_{cll}) - \frac{4}{5}
\lambda_{ll} (\lambda_{a} \lambda_{a}) (\lambda_{bll} \lambda_{bll}) - 2
(\lambda_{a}\lambda_{all})(\lambda_{bc}\lambda_{b}\lambda_{cll})+
\\
& &+\frac{4}{5} \lambda_{ll} ( \lambda_{a} \lambda_{all})^2 + (\lambda_{a} \lambda_{a})
(\lambda_{bc}\lambda_{bll}
\lambda_{cll}) + \\
& &+4\Lambda\lambda^{2}_{ab}\lambda_{all} \lambda_{bll}- 2\Lambda tr \lambda^{2}_{ab}
\lambda_{cll} \lambda_{cll} -\frac{16}{5}\Lambda \lambda_{ll} \lambda_{ab} \lambda_{all}
\lambda_{bll}+
\\
& & + \frac{34}{25}\Lambda\lambda^{2}_{ll}  \lambda_{all} \lambda_{all}+
 16 \Lambda \lambda_{ppll} \left( - \frac{37}{375}\lambda^{3}_{ll} +
\frac{2}{5} \lambda_{ll}(tr \lambda^{2}_{ab}) - \frac{1}{3} tr
\lambda^{3}_{ab} \right) + \\
& & + \frac{4}{75} \lambda_{ll}^2 (tr \lambda_{ab}^3) - \frac{8}{125} \lambda_{ll}^3 (tr
\lambda_{ab}^2) + \frac{4}{15} \cdot \frac{37}{625}\lambda_{ll}^5 \, .
\end{eqnarray*}
It is interesting that $\phi'^{k}$ has been determined except for 4 scalar functions
$H_{0},H_{1}, H_{2}, H_{3}$. Instead of this, if we have used only the representation
theorems without imposing the entropy principle and the Galilean invariance, we would
have obtained that  $\phi'^{k}$ was depending on 6 arbitrary scalar functions, being a
linear combination of $\lambda_{kll}$, $ \lambda_{kh} \lambda_{hll}$,
$\lambda^{2}_{ka}\lambda_{all}$, $\lambda_{k}$, $\lambda_{ka}\lambda_{a}$,
$\lambda^{2}_{ka}\lambda_{a} $. We note that also  $h'$ is determined in terms of
$H_{0}$,$H_{1}$, $H_{2}$, $H_{3}$. These are arbitrary functions of the 8 scalars $X_1$ - $X_8$.\\
Instead of this, if we have used only the representation theorems without imposing the
entropy principle and the Galilean invariance, we would have obtained that all the scalar
function are arbitrary functions of the following 14 scalars $\lambda_{ll}$, $tr
\lambda^{2}_{ab}$, $tr \lambda^{3}_{ab}$, $\lambda_{all} \lambda_{all}$, $\lambda_{all}
\lambda_{a}$ , $\lambda_{a} \lambda_{a}$, $\lambda_{ab} \lambda_{all} \lambda_{bll}$,
$\lambda_{ab}\lambda_{a}\lambda_{bll}$, $\lambda_{ab}\lambda_{a}\lambda_{b}$,
$\lambda^{2}_{ab}\lambda_{all}\lambda_{bll}$, $\lambda^{2}_{ab}\lambda_{a}\lambda_{bll}$,
$\lambda^{2}_{ab}\lambda_{a}\lambda_{b}$, $\lambda_{ppll}$,
$\Lambda$ . \\
It is useful now to verify these results: To this end we can substitute eqs.
(\ref{ange2}), (\ref{ange2biss}) and (\ref{ange3}) into (\ref{mel3}) and obtain that they
are identically satisfied. The corresponding calculations  are long, so it will be useful
to subdivide them with the following steps.
\begin{itemize}
  \item Firstly we can verify that $(\ref{mel3})_1$ is satisfied with $X_i$ instead of $h'$,
  for $i= 1, \cdots , 8$; consequently, for the theorem on derivation of composite functions,
  it will be satisfied by whatever function of $X_i$, as $h'$ is. But we have to note
  that, if we simplify $X_8$ through the  Hamilton-Kayley theorem
\end{itemize}
\begin{eqnarray*}
\lambda^{3}_{ab} = \lambda_{ll} \lambda^{2}_{ab} + \frac{1}{2} \left( tr \lambda^{2}_{cd}
- \lambda^{2}_{ll} \right) \lambda_{ab} + \left( \frac{1}{3} tr \lambda^{3}_{cd} -
\frac{1}{2} \lambda_{ll} tr \lambda^{2}_{cd} + \frac{1}{6} \lambda^{3}_{ll} \right)
\delta_{ab}
\end{eqnarray*}
it will become more complicate to verify that $X_8$ is a solution, because it will be
necessary also to use the identity
\begin{eqnarray*}
0 &=& \delta_{ij} \left( - \lambda^{2}_{ab}\lambda_{all}\lambda_{bll} + \lambda_{ll}
\lambda_{ab}\lambda_{all}\lambda_{bll}+ \frac{1}{2} \lambda_{all}\lambda_{all} tr
\lambda^{2}_{cd} - \frac{1}{2} \lambda_{all}\lambda_{all} \lambda^{2}_{ll} \right)+ \\
&{}& + \lambda_{ij}\left( - \lambda_{ab}\lambda_{all}\lambda_{bll} + \lambda_{ll}
\lambda_{all}\lambda_{all} \right)- \lambda^{2}_{ij} \lambda_{all}\lambda_{all} +
\lambda_{ill}\lambda_{jll} \, \frac{1}{2} ( \lambda^2_{ll}- tr \lambda^2_{cd} )  + \\
&{}& - 2 \lambda_{ll} \lambda_{ll(i}\lambda_{j)b}\lambda_{bll} +
\lambda_{ia}\lambda_{all}\lambda_{jb}\lambda_{bll} + 2
\lambda_{ll(i}\lambda^2_{j)b}\lambda_{bll}
\end{eqnarray*}
which can be easily proved in the reference frame where  $\lambda_{ab}$ has the diagonal
form.
\begin{itemize}
  \item The second step is to verify that $(\ref{mel3})_2$ is satisfied in the case
   $H_0=1$, $H_1=0$, $H_2=0$, $H_3=0$. Similarly for the case $H_0=0$, $H_1=1$,
  $H_2=0$, $H_3=0$; then for the case $H_0=0$, $H_1=0$, $H_2=1$, $H_3=0$ and, lastly, for the case
   $H_0=0$, $H_1=0$, $H_2=0$, $H_3=1$. Consequently, it will be satisfied for all
   constant values of $H_0$, $H_1$, $H_2$, $H_3$. After  that it is satisfied also in
   the general case: In fact, for the property on  derivation of a product, the terms
   where  $H_0$, $H_1$, $H_2$, $H_3$ are not differentiated will simplify, for what above said, and it remains
\end{itemize}
\begin{eqnarray*}
  \sum_{j=0}^3 V^k_j \cdot \sum_{i=1}^8 \frac{\partial H_j}{\partial X_i}
\end{eqnarray*}
for the right hand side of $(\ref{mel3})_1$ written with $X_i$ instead of $h'$; the
result is zero for what already verified in the first step.
\section{The principles of entropy and of Galilean relativity}
We want that our system $(\ref{0.1})$ is symmetric and hyperbolic, with all the
consequent nice mathematical properties. To this end we impose that all the solution of
eqs. $(\ref{0.1})$ satisfy also the entropy inequality
\begin{equation*}
\frac{\partial h}{\partial t}+\frac{\partial \phi_{k}}{\partial x_{k}}= \sigma\geq 0.
\end{equation*}
This is equivalent to assume the existence of Lagrange Multipliers  $\lambda$,
$\lambda_{i}$, $\lambda_{ij}$, $\lambda_{ill}$, $\lambda_{iill}$ such that
\begin{eqnarray}\label{0.3}\nonumber
   dh &=& \lambda dF+\lambda_{i}dF_{i}+\lambda_{ij}dF^{ij}+\lambda_{ill}dF^{ill}+\lambda_{iill}dF^{iill} \, ,
  \\
  d\phi_{k} &=& \lambda
  dF_{k}+\lambda_{i}dG_{ik}+\lambda_{ij}dG_{ijk}+\lambda_{ill}dG_{illk}+\lambda_{iill}dG_{iillk}
\end{eqnarray}
besides a residual inequality which we leave for the sake of brevity. The Lagrange
Multipliers are also called "mean field". Let us now impose the Galilean relativity
principle by considering the following change of independent variables
\begin{eqnarray}\label{0.7} F_{i_1i_2...i_n}&=&\sum_{k=0}^n{n\choose
k}m_{(i_1i_2...i_k}v_{i_{k+1}...i_n)}
\end{eqnarray}
which can be found in  \cite{3}  and that, applied to our case, becomes
\begin{eqnarray}\label{0.8}
F&=&m,\,\\\nonumber \quad F_{i}&=&mv_{i}+m_{i},\\\nonumber \quad F_{ij}&=&mv_{i}v_{j}+m_{ij}+2m_{(i}v_{j)} \, ,\\
\nonumber
  F_{ill}&=&m_{ill}+m_{ll}v_{i}+2m_{il}v_{l}+mv^{2}v_{i}+m_{i}v^2+2m_{l}v_{i}v_{l}\, ,\\
\nonumber
 F_{iill}&=&m_{iill}+mv^{4}+4m_{i}v_{i}v^2+2m_{ii}v^2+4m_{il}v_{i}v_{l}+4m_{iil}v_{l}.
\end{eqnarray}
Also in \cite{3} we can find how change the constitutive functions $G_{ki_{1}\cdots
i_{n}}$ when the reference frame changes, i.e.,
\begin{eqnarray}\label{0.9}
H_{ki_1i_2\cdots i_n}&=&\sum_{j=0}^n{n\choose j}M_{k(i_1\cdots i_j}v_{i_{j+1}\cdots i_n)}
\end{eqnarray}
where the functions $H_{ki_{1}\cdots i_{n}}$ are defined by
\begin{eqnarray}\label{0.2}
G_{ki_{1}\cdots i_{n}}= v_{k} F_{i_{1}\cdots i_{n}}+H_{ki_{1}\cdots i_{n}}.
\end{eqnarray}
It is interesting that  (\ref{0.9}) looks like (\ref{0.7}), except that they don' t act
on the index  $k$. In our particular case, eqs. (\ref{0.9}) become
\begin{eqnarray}\label{0.10}
\nonumber
H_{k}&=&M_{k} \, , \\
\nonumber
H_{ki}&=&M_{k}v_{i}+M_{ki} \, , \\
\nonumber
H_{kij}&=&M_{k}v_{i}v_{j}+2M_{k(i}v_{j)}+M_{kij} \, , \\
\nonumber
H_{kill}&=&M_{k}v_{i}v^2+M_{ki}v^2+2M_{kl}v_{i}v_{l}+2M_{kil}v_{l}+M_{kll}v_{i}+M_{kill} \, , \\
\nonumber
H_{kiill}&=&M_{k}v^4+4M_{ki}v_{i}v^2+2M_{kii}v^2+4M_{kil}v_{l}v_{i}+4M_{kiil}v_{l}+\,\\
&+&M_{kiill}
\end{eqnarray}
Consequently, the functions  $G_{ki_1\cdots i_n}$ transform as follows
\begin{eqnarray}\label{0.11}
G_{ki}&=&mv_{i}v_{k}+m_{i}v_{k}+M_{k}v_{i}+M_{ki} \, , \\
\nonumber
G_{kij}&=&mv_{i}v_{j}v_{k}+m_{ij}v_{k}+2m_{(i}v_{j)}v_{k}+M_{k}v_{i}v_{j}+2M_{k(i}v_{j)}+M_{kij}\, , \\
\nonumber
G_{kill}&=&m_{ill}v_{k}+m_{ll}v_{i}v_{k}+2m_{il}v_{l}v_{k}+mv^{2}v_{i}v_{k}+m_{i}v^2v_{k}+\,\\\nonumber
&+&2m_{l}v_{i}v_{l}v_{k}+M_{ki}v^2+2M_{kl}v_{i}v_{l}+M_{kll}v_{i}+2M_{kil}v_{l}+\,\\\nonumber
&+&M_{k}v_{i}v^2+M_{kill},\,\\\nonumber
G_{kiillkk}&=&m_{iill}v_{k}+mv^{4}v_{k}+4m_{i}v_{i}v^2v_{k}+2m_{ii}v^2v_{k}+4m_{il}v_{i}v_{l}v_{k}+\,\\\nonumber
&+&4m_{iil}v_{l}v_{k}+M_{k}v^4+4M_{ki}v_{i}v^2+2M_{kii}v^2+4M_{kil}v_{i}v_{l}+\,\\\nonumber&+&4M_{kiil}v_{l}+M_{kiill}.
\end{eqnarray}
We note that from (\ref{0.2}) and (\ref{0.9}), for $n=0$, and from $G_k=F_k$ it follows
$M_k=F_k-Fv_k$. This and  $(\ref{0.8})_{1,2}$ yields  $M_k=m_k$. The new variables  $m$,
$m_{i}$, $m_{ij}$, $m_{ill}$, $m_{iill}$ and $M_{i}$, $M_{ij}$, $M_{kij}$, $M_{kill}$,
$M_{kiill}$ have the same simmetries of  $F_{i_1...i_n}$ and
$G_{ki_1...i_n}$.\\
Let us now substitute into eqs.  $(\ref{0.3})$ the expressions which we have above found
for the variables and the constitutive functions. In this way eqs.  $(\ref{0.3})$ become
\begin{eqnarray}\label{0.12}\nonumber
dh &=& \lambda^I
dm+\lambda^I_{i}dm_{i}+\lambda^I_{ij}dm_{ij}+\lambda^I_{ill}dm_{ill}+\,\\\nonumber
&+&\lambda^I_{iill}dm_{iill}+
(\lambda^I_{i}m+2\lambda^I_{ij}m_{j}+\lambda^I_{jll}m_{ll}\delta_{ij}+\,\\
&+&2\lambda^I_{jll}m_{ij}+ 4\lambda^I_{ppqq}m_{ill})dv_{i}
\end{eqnarray}
\begin{eqnarray}\label{0.12a}\nonumber
d(\phi_{k})&=&(\lambda^I dm+\lambda^I_{i} dm_{i}+\lambda^I_{ij} dm_{ij}+\lambda^I_{ill}
dm_{ill}\,\\\nonumber&+&\lambda^I_{iill} dm_{iill})v_{k}+\lambda^I
dM_{k}+\lambda^I_{i}dM_{ki}+\lambda^I_{il} dM_{kil}+\,\\\nonumber&+&\lambda^I_{ill}
dM_{kill}+\lambda^I_{iill} dM_{kiill}+(\lambda^I
m+\lambda^I_{i}m{i}+\lambda^I_{ij}m_{ij}\,\\\nonumber
&+&\lambda^I_{ill}m_{ill}+\lambda^I_{ppqq}m_{iill})dv_{k}+(\lambda^I_{i}M_{k}+2\lambda^I_{ij}M_{kj}\,\\&+&\lambda^I_{ipp}M_{kll}+2\lambda^I_{lpp}M_{kil}+4\lambda^I_{ppqq}M_{kill})dv_{i},
\end{eqnarray}
with
\begin{eqnarray}\label{0.13}\nonumber
\lambda^I&=&\lambda+\lambda_{i}v_{i}+\lambda_{ij}v_{i}v_{j}+\lambda_{ill}v_{i}v^{2}+\lambda_{ppqq}v^{4}\,
,\\ \nonumber
\lambda^I_{i}&=&\lambda_{i}+2\lambda_{ij}v_{j}+\lambda_{ill}v^{2}+2\lambda_{jpp}v_{i}v_{j}+4\lambda_{ppqq}v_{i}v^{2}\,
,\\ \nonumber
\lambda^I_{ij}&=&\lambda_{ij}+\lambda_{hpp}v_{h}\delta_{ij}+2\lambda_{pp(i}v_{j)}+4\lambda_{ppqq}v_{j}v_{i}+2\lambda_{hhpp}v^{2}\delta_{ij}\,
,\\\nonumber \lambda_{ill}&=&\lambda_{ipp}+4\lambda_{hhpp}v_{i} \,
\\\nonumber
\lambda^I_{iill}&=&\lambda_{ppqq}
\end{eqnarray}
The Galilean relativity principle imposes that $h$, $\phi_{k}-hv_{k}$, $M_{ki}$,
$M_{kij}$, $M_{kill}$, $M_i$ don' t depend on  $v_{i}$. For this condition on  $h$ and
$\phi_{k}-hv_{k}$ we obtain
\begin{eqnarray}\label{0.14}
\frac{\partial{h}}{\partial{v_{i}}}=0
&=&m\lambda^I_{i}+2\lambda^I_{ij}m_{j}+\lambda^I_{jpp}(m_{ll}\delta_{ij}+2m_{ij})+\,\nonumber\\&+&4\lambda^I_{ppqq}m_{ill}\nonumber\\
\frac{\partial{(\phi^k-hv_k)}}{\partial{v_{i}}}=0
&=&M_{k}\lambda^I_{i}+2M_{kj}\lambda^I_{ij}+M_{kll}\lambda^I_{ipp}+2M_{kil}\lambda^I_{lpp}+\,\nonumber\\
&+&4\lambda^I_{ppqq}M_{kill}+{h'}\delta_{ik}\,,
\end{eqnarray}
where $h'$ is defined by
\begin{eqnarray*}
  h' &=& \lambda^I
  m+\lambda^I_{i}m_{i}+\lambda^I_{il}m_{il}+\lambda^I_{ill}m_{ill}+\lambda^I_{iill}m_{iill}-h;
  \end{eqnarray*}
In this way eqs.  $(\ref{0.12})$ and $(\ref{0.12a})$ become respectively
\begin{eqnarray}\label{0.15bis}
 dh^I &=& \lambda^I dm +\lambda^I_{i}dm_{i}+\lambda^I_{il}dm_{il}+\lambda^I_{ill}dm_{ill}+\lambda^I_{iill}dm_{iill} \\
  \nonumber
 d\phi_{k}^I &=&\lambda^I
 dM_{k}+\lambda^I_{i}dM_{ki}+\lambda^I_{il}dM_{kil}+\lambda^I_{ill}dM_{kill}+\lambda^I_{ppqq}dM_{kiill}
\end{eqnarray}
with   $\phi_{k}^I=\phi_k-hv_k$. Let us also define
\begin{eqnarray*}
  \phi'_{k} &=&
  \lambda^I
  M_{k}+\lambda^I_{i}M_{ki}+\lambda^I_{il}M_{kil}+\lambda^I_{ill}M_{kill}+\lambda^I_{ppqq}M_{kiill}-\phi^I_{k}\,;
\end{eqnarray*}
so that eqs.  $(\ref{0.15bis})$ become
\begin{eqnarray}\label{0.17a}\nonumber
 dh' &=& md\lambda^I +m_{i}d\lambda^I_{i}+m_{il}d\lambda^I_{il}+m_{ill}d\lambda^I_{ill}+m_{iill}d\lambda^I_{iill} \\
 d\phi_{k}' &=&M_{k}d\lambda^I+M_{ki}d\lambda^I_{i}+M_{kil}d\lambda^I_{il}+M_{kill}d\lambda^I_{ill}+M_{kiill}d\lambda^I_{ppqq}
 \,
 \end{eqnarray}
from which, by taking $\lambda^I$, $\lambda^I_{i}$, $\lambda^I_{ij}$, $\lambda^I_{ill}$,
$\lambda^I_{iill}$ as independent variables and, by taking the derivatives with respect
to the various components of the mean field, it follows
\begin{eqnarray}\label{0.17}\nonumber
m&=&\frac{\partial h'}{\partial\lambda^I},\quad m_{i}=\frac{\partial
h'}{\partial\lambda^I_{i}},\quad m_{il}=\frac{\partial
h'}{\partial\lambda^I_{il}},\,\\\nonumber\quad m_{ill}&=&\frac{\partial
h'}{\partial\lambda^I_{ill}},\quad m_{iill}=\frac{\partial h'}{\partial\lambda^I_{iill}}
\, ,\\~
\\
M_{k}&=&\frac{\partial\phi'_{k}}{\partial\lambda^I},\quad
M_{ki}=\frac{\partial\phi'_{k}}{\partial\lambda^I_{i}},\quad
M_{kil}=\frac{\partial\phi'_k}{\partial \lambda^I_{il}},\,\nonumber
\\\nonumber
M_{kill}&=&\frac{\partial\phi'_{k}}{\partial\lambda^I_{ill}},\quad
M_{kiill}=\frac{\partial\phi'_{k}}{\partial\lambda^I_{iill}}.
\end{eqnarray}
These last equations are noting more than (\ref{mel1}), but in the new reference frame.
By substituting in  (\ref{0.14}) from (\ref{0.17}) we obtain equations whose expression,
in the previous reference frame, are (\ref{mel3}). Consequently, (\ref{mel1}) and
(\ref{mel3}) are equivalent to the principles of entropy and of Galilean relativity.
Other interesting aspects can be found in \cite{4}, which are adapted for the present
case in \cite{5} and \cite{6}. In order to find the general solution of (\ref{mel3}), let
us write a relativistic counterpart of our equations.

\section{Relativistic extended thermodynamics for dense gases and macromolecular fluids}
Let us consider the balance equations
\begin{equation}\label{1}
\partial_{\alpha}T^{\alpha\beta}=0 \quad , \quad
\partial_{\alpha}A^{\alpha\beta\gamma}=I^{\beta\gamma} \, .
\end{equation}
where $T^{\alpha\beta}$ isn' t  symmetric, while $A^{\alpha\beta\gamma}$ and
$I^{\beta\gamma}$ are symmetric only with respect to the indexes  $\beta\gamma$. The
first of these is the conservation law of momentum-energy, while the trace of the second
one is the conservation law of mass, so that
\begin{equation}\label{3}
   I^{\beta\gamma}g_{\beta\gamma}=0 \, .
\end{equation}
The entropy principle is expressed in terms of  $h^{\alpha}$, called (entropy density -
entropy flux density) tensor, such that
\begin{equation}\label{4}\
\partial_{\alpha}h^{\alpha}=\sigma\geq0
\end{equation}
for every solution of eqs.  $(\ref{1})$.\\
For  Liu Theorem   \cite{7} eq. $(\ref{4})$ is equivalent to assuming the existence of
the Lagrange multipliers $\lambda_{\beta}$,\;\;$\lambda_{\beta\gamma}$ such that
\begin{equation}\label{5}
    \partial_{\alpha}h^{\alpha}-\sigma-\lambda_{\beta}\partial_{\alpha}T^{\alpha\beta}-\lambda_{\beta\gamma}(\partial_{\alpha}A^{\alpha\beta\gamma}-I^{\beta\gamma})=0
\end{equation}
for every value of the independent variables. By differentiating it becomes
\begin{eqnarray}\label{6}
dh^{\alpha}=\lambda_{\beta}dT^{\alpha\beta}+\lambda_{\beta\gamma}dA^{\alpha\beta\gamma}
\, ;\quad  -\sigma+\lambda_{\beta\gamma}I^{\beta\gamma}=0  .\,
\end{eqnarray}
If we define $h'^{\alpha}$ by
\begin{equation}\label{7}
 h^{\alpha}= -h'^{\alpha}+\lambda_{\beta}T^{\alpha\beta}+\lambda_{\beta\gamma}A^{\alpha\beta\gamma}
\end{equation}
then eq.  (\ref{6}) can be rewritten as
\begin{equation}\label{9}
   \ dh'^{\alpha}=\ T^{\alpha\beta}d{\lambda}_{\beta}+\ A^{\alpha\beta\gamma}d{\lambda}_{\beta\gamma} .
\end{equation}
This last equation, by taking $\lambda_{\beta}$,\;\;$\lambda_{\beta\gamma}$ as
independent variables, becomes
\begin{equation}\label{10}
T^{\alpha\beta}=\frac{\partial h'^{\alpha}}{\partial \lambda_{\beta}} \quad , \quad
A^{\alpha\beta\gamma}=\frac{\partial h'^{\alpha}}{\partial \lambda_{\beta\gamma}}
\end{equation}
It follows that from the knowledge of   $h'^{\alpha}$ we obtain $T^{\alpha\beta}$ and
$A^{\alpha\beta\gamma}$; we have only a condition on $h'^{\alpha}$: it has to satisfy the
Einsteinian relativity principle. For well known representation theorems as \cite{8} and
\cite{9}, we have that
\begin{equation}\label{11}
    \ h'^{\alpha}=\
    h_{0}\lambda^{\alpha}+ h_{1}\lambda^{\alpha\gamma}\lambda_{\gamma}+ h_{2} \stackrel{2}{\lambda}{} ^{\alpha\gamma} \lambda_{\gamma}+\ h_{3}{\stackrel{3}{\lambda}}{}^{\alpha\gamma}\lambda_{\gamma}
\end{equation}
where the following definitions have been used
\begin{eqnarray*}
  \stackrel{2}{\lambda}{} ^{\alpha\gamma}=\lambda^{\alpha\beta}\lambda_{\beta\delta}g^{\delta\gamma} \quad ,
  \quad \stackrel{3}{\lambda}{} ^{\alpha\gamma}=\lambda^{\alpha\beta}\lambda_{\beta\delta}\lambda^{\delta\gamma}
\end{eqnarray*}
 with $h_{i}$ scalar functions depending on
\begin{eqnarray}\label{13}
&{}& Q_{1} = \lambda^{\beta}_{\beta} \quad , \quad  Q_{2} = \stackrel{2}{\lambda}{}
^{\alpha\gamma}g_{\alpha\gamma} \quad , \quad   Q_{3} = \stackrel{3}{\lambda}{}
^{\alpha\gamma}g_{\alpha\gamma} \quad , \quad  Q_{4} =
\stackrel{4}{\lambda}{} ^{\alpha\gamma}g_{\alpha\gamma}  \quad ,  \\
&{}& \nonumber  P_{0} = \lambda_{\beta}\lambda^{\beta} \quad , \quad   P_{1} =
\lambda_{\beta}\lambda_{\gamma}\lambda^{\beta\gamma}  \quad , \quad    P_{2} =
\lambda_{\beta}\lambda_{\gamma}\stackrel{2}{\lambda}{}^{\beta\gamma} \quad , \quad P_{3}
= \lambda_{\beta}\lambda_{\gamma}\stackrel{3}{\lambda}{} ^{\beta\gamma} \quad .
\end{eqnarray}
\section{The non relativistic limit of the previous model}
In order to take this limit, let us consider a modified procedure of that used  in
\cite{10}, \cite{11} for ideal gases. We assume that this procedure holds also for
macromolecular gases; this assumption doesn't lead to wrong results because we have
already verified, at the end of sect. 1, that these results are correct. Let us subdivide
the procedure in two parts.

\subsection{A first transformation in 3-dimensional form}
Eq. $(\ref{1})_{1}$ for $\beta = 0,i$ becomes
\begin{equation}\label{14}
    \frac{1}{c}\partial_{t}T^{00}+\partial_{k}T^{k0}=0 \, ,
\end{equation}
\begin{equation}\label{15}
    \frac{1}{c}\partial_{t}T^{0i}+\partial_{k}T^{ki}=0 \, .
\end{equation}
Similarly, eq.  $(\ref{1})_{2}$ for $\beta\gamma=00$, $\beta\gamma=0i$, $\beta\gamma=ij$
becomes
\begin{eqnarray}\label{16}
\frac{1}{c}\partial_{t}A^{000}+\partial_{k}A^{k00} = I^{00}\, , \,
\frac{1}{c}\partial_{t}A^{00i}+\partial_{k}A^{k0i} = I^{0i}\, , \,
\frac{1}{c}\partial_{t}A^{0ij}+\partial_{k}A^{kij} = I^{ij} \, .
\end{eqnarray}
Now we adopt the following change of variables
\begin{eqnarray}\label{17}
T^{00}=m^{4}_{0}c F_{2},\
 T^{k0}=m^{4}_{0}G_{2}^{k},\
 T^{0i}=m^{4}_{0}F_{2}^{i},\
T^{ki}=\frac{m^{4}_{0}}{c}G_{2}^{ki}
 \end{eqnarray}
 \begin{eqnarray}\label{18}
 A^{000}=
m^{5}_{0}c^{2}F_{3},\
 A^{00i}=m^{5}_{0}cF^{i}_{3},\
A^{0ij}=m^{5}_{0}F^{ij}_{3}
\end{eqnarray}
\begin{eqnarray}\label{18bis}
    \ A^{k00}=m^{5}_{0}cG^{k}_{3},\
A^{k0i}=m^{5}_{0}G^{ki}_{3},\ A^{kij}= m^{5}_{0}\frac{1}{c}G^{kij}_{3}
\end{eqnarray}
\begin{eqnarray}\label{19}
I^{00}= c R m^{5}_{0},\ I^{0i}=Q^{i} m^{5}_{0},\ I^{ij}= \frac{1}{c}p^{ij}m^{5}_{0} \, .
\end{eqnarray}
In their terms eqs. $(\ref{14})$, $(\ref{15})$ and $(\ref{16})$ become
\begin{eqnarray}\label{20}
&{}& \partial_{t}F_{2}+\partial_{k}G^{k}_{2} = 0 \quad , \quad
 \partial_{t}F^{i}_{2}+\partial_{k}G^{ki}_{2} = 0  \quad , \quad
 \partial_{t}F_{3}+\partial_{k}G^{k}_{3} = R \\
 \nonumber \\
&{}&  \nonumber \partial_{t}F^{i}_{3}+\partial_{k}G^{ki}_{3} = Q^{i} \quad , \quad
 \partial_{t}F^{ij}_{3}+\partial_{k}G^{kij}_{3} = p^{ij} \, .
 \end{eqnarray}
The last of these can be also subdivided in
\begin{equation}\label{20bis}
\partial_{t}F^{ll}_{3}+\partial_{k}G^{kll}_{3} =p^{ll}  \quad , \quad
\partial_{t}F^{<ij>}_{3}+\partial_{k}G^{k<ij>}_{3} =p^{<ij>} \, ,
\end{equation}
while eq. $(\ref{4})$  can be  subdivided in
 \begin{equation}\label{21}
 \frac{1}{c}\partial_{t}h^{0}+\partial_{k}h^{k} = \sigma \, .
 \end{equation}
With the following changes of names
 \begin{equation}\label{22}
 \ h^{0} = m^{3}_{0}h,\;\;\;\;\ h^{k} =
 \frac{m^{3}_{0}}{c}\phi^{\kappa}, \;\;\;\;\ \sigma =
 \frac{m^{3}_{0}}{c}\sigma^{\star}
 \end{equation}
eq. $(\ref{21})$ becomes
 \begin{equation}\label{23}
    \partial_{t}h+\partial_{k}\phi^{k} = \sigma^{\star} \, .
\end{equation}
\subsection{A first transformation of the Lagrange Multipliers}
EQ. $(\ref{6})_{1}$ for $\alpha = 0$ becomes
\begin{equation}\label{24}
    \ dh^{0} =
    \lambda_{0}dT^{00}+\lambda_{i}dT^{0i}+\lambda_{00}dA^{000}+2\lambda_{0i}dA^{00i}+\lambda_{ij}dA^{0ij}
    \, .
\end{equation}
By using eq. $(\ref{22})_{1}$ and $(\ref{17})$-$(\ref{18bis})$,  it transforms into
\begin{eqnarray}\label{25}
    \ d(m^{3}_{0}h) &=&
    \lambda_{0}d(m^{4}_{0}c F_{2})+\lambda_{i}d(m^{4}_{0}F^{i}_{2})+\lambda_{00}d(m^{5}_{0} c^{2} F_{3})+\,\\\nonumber
    &+&
    2\lambda_{0i}d(m^{5}_{0}cF^{i}_{3})+\lambda_{ij}d(m^{5}_{0}F^{ij}_{3})
\end{eqnarray}
from which
\begin{eqnarray}\label{26}
 d h &=&\lambda_{0}m_{0} c
 d(F_{2})+\lambda_{i}m_{0}d(F^{i}_{2})+\lambda_{00}m^{2}_{0}c^{2}d(F_{3})+\,\\\nonumber
 &+& 2\lambda_{0i}m^{2}_{0}cd(F^{i}_{3})+\lambda_{ij}m^{2}_{0}d(F^{ij}_{3})=\,\\\nonumber &=&
 \ell d F_{2}+\ell_{i} d F^{i}_{2}+\eta d F_{3}+\mu_{i}dF^{i}_{3}+\mu_{ij} dF^{ij}_{3} \, ,
\end{eqnarray}
with
\begin{equation}\label{27}
\ell = \lambda_{0}m_{0}c \, , \, \ell_{i} = \lambda_{i}m_{0} \, , \, \eta =
\lambda_{00}m^{2}_{0}c^{2} \, , \, \mu_{i} = 2\lambda_{0i}m^{2}_{0}c \, , \, \mu_{ij} =
\lambda_{ij}m^{2}_{0}\, .
\end{equation}
Similarly, eq. $(\ref{6})_{1}$ for $\alpha=k$ becomes
\begin{equation}\label{28}
   d h^{k}=
    \lambda_{0}dT^{k0}+\lambda_{i}dT^{ki}+\lambda_{00}d A^{k00}+2\lambda_{0i}d
    A^{k0i}+\lambda_{ij}dA^{kij}.
\end{equation}
By using eq.  $(\ref{22})$ and $(\ref{17})$, $(\ref{18bis})$, it transforms into
\begin{eqnarray}\label{29}\nonumber
 d (\frac{m^{3}_{0}}{c}\phi^{k})&=&
\lambda_{0}d(m^{4}_{0}G^{k}_{2})+\lambda_{i}d(\frac{m^{4}_{0}}{c}G^{ki}_{2})+\lambda_{00}d(m^{5}_{0}cG^{k}_{3})+\,\\\nonumber\
&+&2\lambda_{0i}d(m^{5}_{0}G^{ki}_{3})+\lambda_{ij}d(m^{5}_{0}\frac{1}{c}G^{kij}_{3})
\end{eqnarray}
which can be rewritten also as
\begin{eqnarray}\label{30}
    d\phi^{k}&=&
    \lambda_{0}cm_{0}d(G^{k}_{2})+\lambda_{i}m_{0}d(G^{ki}_{2})+\lambda_{00}c^{2}m^{2}_{0}d(G^{k}_{3})+\,\\\nonumber
    &+&2\lambda_{0i}cm^{2}_{0}d(G^{ki}_{3})+\lambda_{ij}m^{2}_{0}d(G^{kij}_{3})=\,\\\nonumber &=&
    \ell dG^{k}_{2}+\ell_{i}d G^{ki}_{2}+\eta
    dG^{k}_{3}+\mu_{i}dG^{ki}_{3}+\mu_{ij}dG_{3}^{kij} \, ,
\end{eqnarray}
where we have used  $(\ref{27})$. The equations $(\ref{26})$ and $(\ref{30})$ represent
the entropy principle for the system $(\ref{20})$; moreover, eq. $(\ref{27})$ gives the transformation
of the  Lagrange multipliers.\\
Let us now find the transformation of $h'^{\alpha}$ ; eq. $(\ref{7})$ for $\alpha=0$ is
\begin{equation}\label{31}
 h^{0}=
    -h'^{0}+\lambda_{0}T^{00}+\lambda_{i}T^{0i}+\lambda_{00}A^{000}+2\lambda_{i0}A^{0i0}+\lambda_{ij}A^{0ij}
    \, .
\end{equation}
By using eqs.  $(\ref{22})_{1}$,   $(\ref{27})$ and $(\ref{17})$-$(\ref{18bis})$, eq.
$(\ref{31})$ becomes
\begin{equation}\label{32}
    m^{3}_{0}h = -h'^{0}+ \ell
    m^{3}_{0}F_{2}+\ell_{i}m^{3}_{0}F^{i}_{2}+\eta
    m^{3}_{0}F_{3}+\mu_{i}m^{3}_{0}F^{i}_{3}+\mu_{ij}m^{3}_{0}F^{ij}_{3}.
\end{equation}
This can be rewritten as
\begin{equation}\label{33}
  h =\frac{-h'^{0}}{m^{3}_{0}}+\ell F_{2}+\ell_{i}F^{i}_{2}+\eta
    F_{3}+\mu_{i}F^{i}_{3}+\mu_{ij}F^{ij}_{3}
\end{equation}
or
\begin{equation}\label{34}
     h = -h'+\ell F_{2}+\ell_{i}F^{i}_{2}+\eta
    F_{3}+\mu_{i}F^{i}_{3}+\mu_{ij}F^{ij}_{3}
\end{equation}
where we have defined
\begin{equation}\label{34bis}
    h'^{0} = m^{3}_{0}h' \, .
\end{equation}
In this way we have found the counterpart of $(\ref{22})_{1}$ for $h'^{0}$. Let us now
find the transformation of  $h'^{\alpha}$ for $\alpha=k$; from eq. $(\ref{7})$ we find
\begin{equation}\label{35}
    h^{k} =
    -h'^{k}+\lambda_{0}T^{k0}+\lambda_{i}T^{ki}+\lambda_{00}A^{k00}+2\lambda_{i0}A^{ki0}+\lambda_{ij}A^{kij}
\end{equation}
This, by using again eqs. $(\ref{22})_{2}$,  $(\ref{27})$ and
$(\ref{17})$-$(\ref{18bis})$, can be written as
\begin{eqnarray}\label{36}\nonumber
\frac{m^3_0}{c} \phi^{k}
    &=&
    -\ h'^{k}+\frac{\ell}{m_{0}c}m^{4}_{0}G^{k}_{2}+\frac{\ell_{i}}{m_{0}}\frac{m^{4}_{0}}{c}G^{ki}_{2}+\,\\\nonumber
    &+&\frac{\eta }{m^{2}_{0}c^{2}}m^{5}_{0}c G^{k}_{3}
   +\frac{\mu_{i}}{m^{2}_{0}c}m^{5}_{0}G^{ki}_{3}+\frac{\mu_{ij}}{m^{2}_{0}}\frac{m^{5}_{0}}{c}G^{kij}_{3}
\end{eqnarray}
or
\begin{equation}\label{37}
     \phi^{k} = - \phi'^{k}+\ell G^{k}_{2}+\ell_{i}G^{ki}_{2}+\eta
    G^{k}_{3}+\mu_{i}G^{ki}_{3}+\mu_{ij}G^{kij}_{3}
\end{equation}
with
\begin{equation}\label{37bis}
h'^{k}=\frac{m_{0}^{3}}{c}\phi'^{k} .
\end{equation}
Eqs.  $(\ref{34bis})$  and $(\ref{37bis})$  are the counterparts of  $(\ref{22})_{1,2}$
for $h'^{0}$ and $h'^{k}$. Eqs. $(\ref{34})$ and $(\ref{37})$  are the counterpart of eq.
$(\ref{7})$ for the  system $(\ref{20})$. Let us finish by considering the counterpart of
mass  conservation  $(\ref{3})$, that is $-I^{00} + I^{ij}\delta_{ij}= 0$ or, by use of
$(\ref{19})$,  $- \ c R m^{5}_{0}+ \frac{1}{c}p^{ij}m^{5}_{0}\delta_{ij} = 0$. In other
words,
\begin{equation}\label{39}
R = \frac{1}{c^{2}}p^{ij}\delta_{ij} \, .
\end{equation}

\subsection{The second transformation in 3-dimensional form}
\subsubsection{Suggestions from the kinetic theory for ideal gases}
For ideal gases the variables $F_{2}, F_{2}^{i}, F_{3}, F_{3}^{i}, F_{3}^{ij}$ have
counterparts in statistical mechanics where they are defined as moments of the
distribution function $\widetilde{f}$, i.e., by means of the following integrals
\begin{eqnarray}\label{42}
 F_{2} &=&\int \tilde{f}\gamma^{6}d\mathbf u  \, , \quad
 F_{2}^{i} =\int\tilde{f}\gamma^{6}u^{i} d\mathbf u \, , \\
 F_{3} &=&\int\tilde{f}\gamma^{7}d\mathbf u \nonumber \, , \quad
 F_{3}^{i} =\int\tilde{f} \gamma^{7}u^{i}d\mathbf u \, , \quad
 F_{3}^{ij} =\int\tilde{f}\gamma^{7}u^{i}u^{j}d\mathbf u \, .
 \end{eqnarray}
where $\gamma$ is the Lorentz factor
 $(1-\frac{u^{2}}{c^{2}})^{-\frac{1}{2}}$.\\
If we take the limits of these expressions for $c\rightarrow\infty$ we obtain
$F_{2}=F_{3}$, $F_{2}^{i}=F_{3}^{i}$ so that they will be no more independent variables;
to avoid this problem, we take  suitable invertible linear combinations of the equations
before taking the limits.\\
In particular, as first equation we take the linear combination of $(\ref{20})_{3}$ and $
(\ref{20bis})_{1}$ trough the coefficients 1 and $\frac{-1}{c^{2}}$, respectively; so we
obtain the following eq. $(\ref{59})_{1}$ which is the \textit{conservation law of mass}
with
\begin{eqnarray}\label{44}
F= F_{3}-\frac{1}{c^{2}}F_{3}^{ll}  \, , \quad G^{k}=G^{k}_{3}-\frac{1}{c^{2}}G_{3}^{kll}
\, ;
\end{eqnarray}
moreover, we have taken into account eq.(\ref{39}). \\
As  second equation we take $(\ref{20})_{2}$ which can be written as the following eq.
$(\ref{59})_{2}$ which is the \textit{conservation law of momentum} with
\begin{eqnarray}\label{46}
F^{i} = F^{i}_{2}  \, , \quad G^{ki} = G_{2}^{ki} \, ;
\end{eqnarray}
As third equation we take the linear combination of $(\ref{20})_{1}$, $(\ref{20})_{3}$, $
(\ref{20bis})_{1}$ trough the coefficients $2c^{2}$, $-2c^{2}$, $2$ respectively; so it
becomes
 \begin{equation}\label{47}
    \partial_{t}F^{ll}+\partial_{k}G^{kll} = 0
\end{equation}
which is the \textit{conservation law of energy}, with
\begin{eqnarray}\label{48}
F^{ll} =2c^{2}(F_{2}-F_{3}+\frac{1}{c^{2}}F_{3}^{ll})  \, , \quad G^{kll} =
2c^{2}(G_{2}^{k}-G_{3}^{k}+\frac{1}{c^{2}}G_{3}^{kll}) \, .
\end{eqnarray}
As fourth equation we take $(\ref{20bis})_{2}$ which we write as
\begin{equation}\label{49}
    \partial_{t}F^{<ij>}+\partial_{k}G^{k<ij>} = p^{<ij>}
\end{equation}
where
\begin{equation}\label{50}
    F^{<ij>}=F_{3}^{<ij>}  \, , \quad
    G^{k<ij>}= G^{k<ij>}_{3} \, .
\end{equation}
We transform furtherly eq. $(\ref{49})$ adding to it eq. $(\ref{47})$ multiplied by
$\frac{\delta^{ij}}{3}$ and obtaining the following eq. $(\ref{59})_{3}$ with
\begin{equation}\label{53}
    F^{ij}=F_{3}^{ij}+\frac{c^{2}\delta^{ij}}{3}(2F_{2}-2F_{3}+\frac{F_{3}^{ll}}{c^{2}})
\end{equation}
\begin{equation}\label{54}
    G^{kij}=G^{kij}_{3}+\frac{c^{2}\delta^{ij}}{3}(2G_{2}^{k}-2G_{3}^{k}+\frac{G_{3}^{kll}}{c^{2}})
\end{equation}
Equation  $(\ref{59})_3$ encloses both  $(\ref{47})$ and $(\ref{49})$. As other equation
we take the linear combination of $(\ref{20})_{4}$ and $ (\ref{20})_{2}$ trough the
coefficients $2c^{2}$ and $-2c^{2}$, respectively; so we obtain the following eq.
$(\ref{59})_{4}$ with
\begin{eqnarray}\label{56}
F^{ill}= 2c^{2}(F_{3}^{i}-F_{2}^{i}) \, , \, G^{kill}= 2c^{2}(G_{3}^{ki}-G_{2}^{ki})  \,
, \, p^{ill}=2c^{2}Q^{i} \, .
\end{eqnarray}
Finally, as last equation we take the linear combination of $(\ref{20})_{1}$,
$(\ref{20})_{3}$, $(\ref{20bis})_{1}$ trough the coefficients $-8c^{4}$, $+ 8c^{4}$,
$-4c^{2}$ respectively; so we obtain the following eq. $(\ref{59})_{5}$ with
\begin{eqnarray}\label{58}\begin{array}{c}
                                   F^{iill} =
                                   -8c^{4}F_{2}+8c^{4}F_{3}-4c^{2}F_{3}^{ll}\\
                                   \\G^{kiill} =
                                   -8c^{4}G_{2}^{k}+8c^{4}G_{3}^{k}-4c^{2}G_{3}^{kll}\\
                                   \\p^{iill} =
                                   8c^{4}R-4c^{2}p^{ll}=8c^{2}p^{ll}-4c^{2}p^{ll}=4c^{2}p^{ll}
                                   \, ,
\end{array}
\end{eqnarray}
where eq. (\ref{39}) has been used. The complete system is
\begin{eqnarray}\label{59}
                             &{}& \partial_{t} F+ \partial_{k}G^{k}= 0 \, , \,
                              \partial_{t} F^{i}+ \partial_{k}G^{ki} = 0 \, , \,
                              \partial_{t} F^{ij}+ \partial_{k}G^{kij} = p^{<ij>} \, , \\
                             &{}&  \partial_{t} F^{ill}+ \partial_{k}G^{kill} = p^{ill} \, , \,
                              \partial_{t }F^{iill}+ \partial_{k}G^{kiill} = p^{iill} \,
                              . \nonumber
\end{eqnarray}
Note that the only difference between this system and $(\ref{0.1})$ is that  in
$(\ref{59})_1$ intervenes $G^k$, while in $(\ref{0.1})_1$ there is $F^k$; for this reason
we will have to impose the further condition $G^k=F^k$.
\subsubsection{Reasons for the above choice of coefficients. }
The coefficients in the above linear combinations have been chosen for the following
reasons: from (\ref{42}), $(\ref{44})_1$, $(\ref{46})_1$, $(\ref{48})_1$, $(\ref{50})_1$,
$(\ref{56})_1$, $(\ref{58})_1$ it follows
\begin{eqnarray*}
&{}& F=\int\widetilde{f}\gamma^{7}(1-\frac{u^{2}}{c^{2}})d\mathbf \quad , \quad
F^{i}=\int\widetilde{f}\gamma^{6}u^{i}d\mathbf u u=\int\widetilde{f}\gamma^{5}d\mathbf u
\quad , \quad
F^{ll}=2c^{2}\int\widetilde{f}\gamma^{7}(\frac{1}{\gamma}-1+\frac{u^{2}}{c^{2}})d\mathbf
u \quad , \\
&{}& F^{<ij>}=\int\widetilde{f}\gamma^{7}(u^{i}u^{j}-\frac{1}{3}u^{2}\delta^{ij})d\mathbf
u \quad , \quad
F^{ill}=2c^{2}\int\widetilde{f}\gamma^{7}u^{i}(1-\frac{1}{\gamma})d\mathbf u \quad , \\
&{}&
F^{iill}=\int\widetilde{f}\gamma^{7}(-8c^{4}\frac{1}{\gamma}+8c^{4}-4c^{2}u^{2})d\mathbf
u \, ,
\end{eqnarray*}
which have limits $\int\widetilde{f}d\mathbf u$,
    $\int\widetilde{f}u^{i}d\mathbf u$,
$\int\widetilde{f}u^{2}d\mathbf u$, $\int\widetilde{f}u^{<i}u^{j>}d\mathbf u$,
$\int\widetilde{f}u^{i}u^{2} d\mathbf u$, $\int\widetilde{f}u^{4}d\mathbf u$, where we
have taken into account also that
\begin{equation}\
    \frac{1}{\gamma}=\sqrt{1-\frac{u^{2}}{c^{2}}}=1-\frac{1}{2} \frac{u^{2}}{c^{2}}-
    \frac{1}{8} \frac{u^{4}}{c^{4}}+\frac{u^{6}}{c^{6}} (...) \, .
\end{equation}
From the previous expressions it follows that $F^{ij}$ has limit $\int\widetilde{f}
u^{i}u^{j}d\mathbf u$. Obviously, the previous properties hold in the case of ideal
gases; we assume that the corresponding change of equations is appropriate also for dense
gases and for macromolecular fluids. In this way generality is not lost because the
Galilean relativity principle has the same form for both cases.
\subsection{Second Transformation of the Lagrange multipliers}
The change of equations in the previous section induces another one on the Lagrange
multipliers and we want now to determine it.\\
We observe that the equations  $(\ref{44})_{1}$, $(\ref{46})_{1}$, $(\ref{53})$,
$(\ref{56})_{1}$, $(\ref{58})_{1}$ give $F$, $F^{i}$, $F^{ij}$, $F^{ill}$, $F^{iill}$ in
terms of $F_{2}$, $F^{i}_{2}$, $F_{3}$, $F_{3}^{i}$, $F^{ij}_{3}$. Let us now take the
inverse of these relations,  which are
\begin{eqnarray}\label{60}
&{}&  F_{2} = \frac{F^{ll}}{2c^{2}}+F \quad , \quad
                             F_{2}^{i} = F^{i}\\
&{}&  F_{3} =
                             F+\frac{F^{ppll}}{4c^{4}}+\frac{F^{ll}}{c^{2}} \quad , \quad
                             F_{3}^{i} = \frac{F^{ill}}{2c^{2}}+F^{i} \quad , \quad
                             F_{3}^{ij} =
                             F^{ij}+\frac{\delta^{ij}}{3}\frac{F^{ppll}}{4c^{2}}\nonumber
\end{eqnarray}
and these we now substitute in eq.  $(\ref{26})$; so it becomes
\begin{eqnarray}\label{61}\nonumber
d h &=& l d(\frac{F^{ll}}{2c^{2}}+F)+l_{i}d(F^{i})+\eta
d(F+\frac{F^{ppll}}{4c^{4}}+\frac{F^{ll}}{c^{2}})+\,\\\nonumber
&+&\mu_{i}d(\frac{F^{ill}}{2c^{2}}+F^{i})+\mu_{ij}d(F^{ij}+\frac{1}{3}\delta^{ij}\frac{F^{ppll}}{4c^{2}})=\,\\
&=& \lambda d
F+\lambda_{i}dF^{i}+\lambda_{ij}dF^{ij}+\lambda_{ill}dF^{ill}\lambda_{ppll}dF^{ppll}
\end{eqnarray}
with
\begin{eqnarray}\label{62}
&{}& \lambda = l+\eta \quad , \quad
                             \lambda_{i} =l_{i}+\mu_{i} \quad , \quad
                             \lambda_{ij}=\mu_{ij}+(\frac{l}{2c^{2}}+\frac{\eta}{c^{2}})\delta_{ij} \quad , \\
&{}& \lambda_{ill}= \frac{\mu_{i}}{2c^{2}} \quad , \quad
                             \lambda_{ppll}=\frac{\eta}{4c^{4}}+\frac{\mu_{ll}}{12c^{2}}\,
                             . \nonumber
\end{eqnarray}
In this way we have found the first part of the entropy principle for the new
system.\\
For the sequel it will be useful to take the inverse of eqs. $(\ref{62})$. They are
\begin{eqnarray}\label{62b}
&{}& \eta =
                              8c^{4}\lambda_{ppll}-\frac{2}{3}c^{2}\lambda_{ll}+\lambda \quad , \quad
                              \ell =
                              -8c^{4}\lambda_{ppll}+\frac{2}{3}c^{2}\lambda_{ll}  \quad , \quad
                              \mu_{i}= 2 c^{2}\lambda_{ill} \quad , \\
&{}& \ell_{i}= \lambda_{i}-2
                              c^{2}\lambda_{ill} \quad , \quad
                              \mu_{ij}=
                              \lambda_{ij}-(4c^{2}\lambda_{ppll}-
                              \frac{\lambda_{ll}}{3}+\frac{\lambda}{c^{2}})\delta_{ij} \quad
                              . \nonumber
\end{eqnarray}
Similarly, eqs. $(\ref{44})_{2}$, $(\ref{46})_{2}$, $(\ref{54})$, $(\ref{56})_{2}$,
$(\ref{58})_{2}$ give $G^{k}$, $G^{ki}$, $G^{kij}$,$G^{kill}$, $G^{kiill}$, in terms of
$G^{k}_{2}$, $G_{2}^{ki}$, $G_{3}^{k}$, $G_{3}^{ki}$ $G_{3}^{kij}$. The inverses of these
relations are
\begin{eqnarray}\label{63}
&{}& G_{2}^{k}=\frac{G^{kll}}{2c^{2}}+G^{k} \quad , \quad
                             G_{2}^{ki}= G^{ki}\\
&{}& G_{3}^{k}=G^{k}+\frac{G^{kppll}}{4c^{4}}+\frac{G^{kll}}{c^{2}} \quad , \quad
                             G_{3}^{ki}=\frac{G^{kill}}{2c^{2}}+G^{ki} \quad , \quad
                             G_{3}^{kij}=G^{kij}+\frac{1}{3}\delta^{ij}\frac{G^{kppll}}{4c^{2}} \quad ,
                             \nonumber
\end{eqnarray}
which now we substitute into eq. $(\ref{30})$  which now becomes
\begin{eqnarray}\label{64}
    d\phi^{k}&=&\ell d(\frac{G^{kll}}{2c^{2}}+G^{k})+\ell_{i}d
    G^{ki}+\eta
    d(G^{k}+\frac{G^{kppll}}{4c^{4}}+\frac{G^{kll}}{c^{2}})+\,\\\nonumber
    &+&\mu_{i}d(\frac{G^{kill}}{2c^{2}}+G^{ki})+\mu_{ij}d(G^{kij}+\frac{1}{3}\delta^{ij}\frac{G^{kppll}}{4c^{2}})=\,\\\nonumber&=&
    \lambda
    dG^{k}+\lambda_{i}dG^{ki}+\lambda_{ij}dG^{kij}+\lambda_{ill}dG^{kill}+\lambda_{ppll}dG^{kppll}\,
    ,
\end{eqnarray}
thanks to eq. (\ref{62}). \\
In this way we have obtained the second part of the entropy principle. We deduce now the
transformation of $ h'$ and $\phi'^{k}$; to this end, let us take eq. $(\ref{34})$ and
substitute eqs. $(\ref{60})$; so we obtain
\begin{eqnarray}\label{65}
    h &=& -h'+l(\frac{F^{ll}}{2c^{2}}+F)+l_{i}F^{i}+\eta(F+\frac{F^{ppll}}{4c^{4}}+\frac{F^{ll}}{c^{2}})+\,\\\nonumber
    &+&\mu_{i}(\frac{F^{ill}}{2c^{2}}+F^{i})+\mu_{ij}(F^{ij}+\frac{\delta^{ij}}{3}\frac
    {F^{ppll}}{4c^{2}})
\end{eqnarray}
which, for eq. (\ref{62}), can be written in the following way
\begin{equation}\label{66}
    h = -h'+\lambda
    F+\lambda_{i}F^{i}+\lambda_{ij}F^{ij}+\lambda_{ill}{F^{ill}}+\lambda_{ppll}{F^{ppll}}
\end{equation}
From eqs.  $(\ref{37})$, by substituting from eqs. $(\ref{63})$ we find
\begin{eqnarray}\label{67}
 \phi^{k}&=&
 -\phi'^{k}+\ell(\frac{G^{kll}}{2c^{2}}+G^{k})+\ell_{i}G^{ki}+\,\\\nonumber
 &+&\eta(G^{k}+\frac{G^{kppll}}{4c^{4}}+\frac{G^{kll}}{c^{2}})+\mu_{i}(\frac{G^{kill}}{2c^{2}}+G^{ki})+\mu_{ij}(G^{kij}+\frac{\delta^{ij}G^{kppll}}{12c^{2}})=\,\\\nonumber
 &=&
 -\phi'^{k}+\lambda
 G^{k}+\lambda_{i}G^{ki}+\lambda_{ij}G^{kij}+\lambda_{ill}G^{kill}+\lambda_{ppll}G^{kppll}\,
 ,
\end{eqnarray}
for eqs.  (\ref{62}). \\
The equations  $(\ref{61})$ and $(\ref{64})$ gives the entropy principle for the balance
equations $(\ref{59})$. The equations $(\ref{66})$ and $(\ref{67})$ give the counterparts
of  $(\ref{34})$ and $(\ref{37})$ for the new system  $(\ref{59})$. We have now to obtain
$h'$ and $\phi'^{k}$ from $(\ref{11})$ expressing them in terms of the new Lagrange
multipliers and then taking the limits for $c\longrightarrow\infty$. The functions $h'$
and $\phi'^{k}$ are called  "potentials".

\section{Determination of the potentials}
With eqs. $(\ref{66})$ and $(\ref{67})$ we have found that the functions  $h'$ and
$\phi'^{k}$ for the new system $(\ref{59})$ are exactly the same obtained, through the
above mentioned passages, from the  4-potential $h'^{\alpha}$ for the  system
$(\ref{1})$. Consequently, we may obtain them from the result $(\ref{11})$ of the
relativistic system. To this end it is necessary to write the relativistic Lagrange
multipliers in terms of those for the  system $(\ref{59})$. By deducing $\lambda_{0}$,
$\lambda_{i}$, $\lambda_{00}$, $\lambda_{0i}$ and $\lambda_{ij}$ from $(\ref{27})$, and
by substituting in their expressions eqs.  $(\ref{62b})$ we obtain
\begin{eqnarray}\label{Mel10}
{\lambda^\beta}_\gamma&=& \frac{c^2}{m_0^2} \left[
\begin{pmatrix}
  -8 \lambda_{ppll} & 0_{j} \\
  {} & {} \\
  0_{i} & -4 \lambda_{ppll} \delta_{ij}
\end{pmatrix} + \frac{1}{c} \begin{pmatrix}
  0 & - \lambda_{jll} \\
    {} & {} \\
  \lambda_{ill} &  0_{ij}
\end{pmatrix} +  \right. \\
&{}& \nonumber \\
&+& \frac{1}{c^2} \left. \begin{pmatrix}
  \frac{2}{3} \lambda_{ll} & 0_{j} \\
    {} & {} \\
  0_{i} & \lambda_{ij} + \frac{1}{3} \lambda_{ll} \delta_{ij}
\end{pmatrix} + \frac{1}{c^4}  \begin{pmatrix}
  - \lambda & 0_{j} \\
    {} & {} \\
  0_{i} & - \lambda \delta_{ij}
\end{pmatrix} \right]  \, , \nonumber \\
&{}& \nonumber \\
{\lambda^\beta}&=& \frac{c^3}{m_0} \left[
\begin{pmatrix}
  8 \lambda_{ppll}  \\
  {}  \\
  0_{i}
\end{pmatrix} + \frac{1}{c} \begin{pmatrix}
  0  \\
    {} \\
  -2 \lambda_{ill}
\end{pmatrix} + \frac{1}{c^2}  \begin{pmatrix}
  - \frac{2}{3} \lambda_{ll}  \\
    {}  \\
  0_{i}
\end{pmatrix} + \frac{1}{c^3}  \begin{pmatrix}
 0 \\
    {}\\
  \lambda_{i}
\end{pmatrix} \right]  \, . \nonumber
\end{eqnarray}
We can now begin to evaluate the 4-vectors intervening  in (\ref{11}) and the scalars
(\ref{13}).
\subsection{The scalars Q$_1$ - Q$_4$ .}
Let us begin with $(\ref{13})_1$. From eq.  $(\ref{Mel10})_1$ we find
\begin{eqnarray}\label{Mel11}
Q_1 = \frac{c^2}{m_0^2} \left( - 20 \lambda_{ppll} + \frac{1}{c^2}\frac{8}{3}
\lambda_{ll} - \frac{4}{c^4}  \lambda \right) \, .
\end{eqnarray}
Consequently, if we assume that the scalar functions depend on  $Q_1$ as composite
functions through  $\frac{m^2_0}{-20 c^2}Q_1$, then their limits will be  functions of
$\lambda_{ppll}=X_1$. \\
Let us consider now eq.  $(\ref{13})_2$. From eq. $(\ref{Mel10})_1$ we find
\begin{eqnarray*}
Q_2 = \frac{c^4}{m_0^4} \left( 112 \lambda_{ppll}^2 + 0(\frac{1}{c}) \right) \, ,
\end{eqnarray*}
where  $0(\frac{1}{c})$ has limit zero when  $c$ goes to infinity. Therefore, if we
assume that the scalar functions depend on  $Q_2$ as composite functions through
$\frac{m^4_0}{c^4}Q_2$, then their limits will be  functions of $112 \lambda_{ppll}^2$.
But this is not independent from $X_1$, so that this result is too much restrictive. The
idea is now to find a number $k$ such that $\frac{m^4_0}{c^4}(Q_2 + k Q_1^2)$ has zero
limit for $c$ going to infinity. We find $k= - \frac{7}{25}$. After that, we see that
\begin{eqnarray}\label{Mel12}
Q_2 - \frac{7}{25} Q_1^2 = \frac{c^2}{m_0^4} \left( - 2 \lambda_{all} \lambda_{all} +
\frac{16}{5}\lambda_{ppll}\lambda_{ll} + 0(\frac{1}{c}) \right) \, .
\end{eqnarray}
Therefore, if we assume that the scalar functions depend on  $Q_1$  and $Q_2$ as
composite functions through $\frac{m^2_0}{-20 c^2}Q_1$ and $\frac{m^4_0}{c^2}(Q_2 -
\frac{7}{25} Q_1^2)$, then their limits will be  functions of $X_1$ ed $X_2$. \\
Going on in a similar way, we look for the numbers $k_1$ and $k_2$ such that
$\frac{m^6_0}{c^4}[Q_3 + k_1 Q_1(Q_2 - \frac{7}{25} Q_1^2) + k_2 Q_1^3]$ has zero limit
for $c$ going to infinity. We find $k_1= - \frac{9}{10}$ and $k_2= - \frac{11}{125}$.
After that, we obtain
\begin{eqnarray}\label{Mel13}
Q_3 - \frac{9}{10} Q_1(Q_2 - \frac{7}{25} Q_1^2) - \frac{11}{125} Q_1^3 =
\frac{c^2}{m_0^6} \left( - \frac{3}{2}X_3 + 0(\frac{1}{c}) \right) \, .
\end{eqnarray}
Then the  limit of a scalar function depends on $X_1$, $X_2$ and $X_3$. \\
Similarly, we search the numbers $k_3$, $k_4$,  $k_5$, $k_6$  such that
$\frac{m^8_0}{c^4}\{Q_4 + k_3 Q_1 [Q_3 - \frac{9}{10} Q_1(Q_2 - \frac{7}{25} Q_1^2) -
\frac{11}{125} Q_1^3] + k_4 (Q_2 - \frac{7}{25} Q_1^2)^2 + k_5 Q_1^2 (Q_2 - \frac{7}{25}
Q_1^2) + k_6 Q_1^4]$ has zero limit for $c$ going to infinity. We find $k_3= -
\frac{16}{15}$, $k_4= - \frac{1}{2}$, $k_5= - \frac{14}{25}$, $k_6= - \frac{19}{625}$.
After that, we obtain
\begin{eqnarray}\label{Mel14}
&{}& Q_4 - \frac{16}{15} Q_1 \left[ Q_3 - \frac{9}{10} Q_1 \left( Q_2 - \frac{7}{25}
Q_1^2 \right) -
\frac{11}{125} Q_1^3 \right] - \frac{1}{2} \left( Q_2 - \frac{7}{25} Q_1^2 \right)^2 \\
&{}&  - \frac{14}{25} Q_1^2 \left( Q_2 - \frac{7}{25} Q_1^2 \right) - \frac{19}{625}
Q_1^4    = \frac{c^2}{m_0^8} \left( - 2X_4 + 0(\frac{1}{c}) \right) \, . \nonumber
\end{eqnarray}
Therefore, the limit of a scalar function depends on  $X_1$, $X_2$, $X_3$ and $X_4$.  \\
Before proceeding with the other scalars $(\ref{13})_{5-8}$, let us evaluate the
4-vectors of which (\ref{11}) is a linear combination.
\subsection{Limits of the 4-vectors in the expression of the relativistic  potential}
From the eqs. (\ref{34bis}), (\ref{37bis}) and $(\ref{Mel10})_2$ it follows that the term
$h_{0}\lambda^{\alpha}$ contributes to $\phi'^{k}$ the term $-2H_{0}\lambda_{kll}$ and to
$h'$ the term $8 H_{0}\lambda_{ppll}$, where $H_0$ is the
limit of $\frac{c^3}{m_0^4}h_0$ for $c$ going to infinity. \\
But, from  eqs. (\ref{37bis}) and $(\ref{Mel10})$ it follows that the term
$h_{1}\lambda^{\alpha\gamma}\lambda_{\gamma}$ contributes to $\phi'^{k}$ the term
$\frac{c^5}{m_0^6}h_1 \, 16 \lambda_{ppll} \lambda_{kll}  $ which is  parallel to the
previous one. In order not to lose generality, it is better to look for a number $a$ such
that $ \left( \lambda^{\alpha\gamma}\lambda_{\gamma} + a \, Q_1 \, \lambda^{\alpha}
\right) \frac{m_0^3}{c^5}$ has zero limit. We find $a= - \frac{2}{5}$. After that, we
obtain
\begin{eqnarray}\label{Mel15}
\lambda^{\alpha\gamma}\lambda_{\gamma} - \frac{2}{5} \,  Q_1 \, \lambda^{\alpha} =
\frac{c^3}{m_0^3} \left[
\begin{pmatrix}
 X_2  \\
  {}  \\
  0_{i}
\end{pmatrix}
+ \frac{1}{c} \begin{pmatrix}
  0  \\
    {} \\
  V_1^i
\end{pmatrix} + 0 \left( \frac{1}{c} \right) \right]
\end{eqnarray}
Now a linear combination, with arbitrary coefficients, of $\lambda^{\alpha}$,
$\lambda^{\alpha\gamma}\lambda_{\gamma}$,
 $\stackrel{2}{\lambda}{} ^{\alpha\gamma} \lambda_{\gamma}$,
 ${\stackrel{3}{\lambda}}{}^{\alpha\gamma}\lambda_{\gamma}$ is also a linear combination, with arbitrary
 coefficients, of
 $\lambda^{\alpha}$, $\lambda^{\alpha\gamma}\lambda_{\gamma} - \frac{2}{5} \,  Q_1 \, \lambda^{\alpha}$,
 $\stackrel{2}{\lambda}{} ^{\alpha\gamma} \lambda_{\gamma}$,
 ${\stackrel{3}{\lambda}}{}^{\alpha\gamma}\lambda_{\gamma}$
 so we can suppose that  in (\ref{11}) there is  $\lambda^{\alpha\gamma}\lambda_{\gamma} - \frac{2}{5} \,  Q_1 \, \lambda^{\alpha}$
 instead of $\lambda^{\alpha\gamma}\lambda_{\gamma}$; After that,
 from  eqs. (\ref{34bis}) and
(\ref{37bis}) it follows that the term $h_{1}(\lambda^{\alpha\gamma}\lambda_{\gamma} -
\frac{2}{5} \, Q_1 \, \lambda^{\alpha})$ contributes to $\phi'^{k}$ the term
$H_{1}(-2\lambda_{ih}\lambda_{hll}+4\lambda_{ppll}\lambda_{i}+\frac{4}{5}\lambda_{ll}\lambda_{ill})$
and to  $h'$ the terms $H_{1}X_2$, where $H_1$ is the
limit of $\frac{c^3}{m_0^6}h_1$ for $c$ going to infinity. \\
Proceeding furtherly  in this way, we search the numbers  $a_1$, $a_2$, $a_3$, such that \\
$\frac{m_0^5}{c^5} \left\{ \stackrel{2}{\lambda}{} ^{\alpha\gamma} \lambda_{\gamma}+ a_1
Q_1 \lambda^{\alpha\gamma}\lambda_{\gamma} + \left[ a_2Q_1^2+ a_3  (Q_2 - \frac{7}{25}
Q_1^2) \right] \lambda^{\alpha} \right\} $, has zero limit for  $c$ going to infinity. We
find  $a_1= - \frac{3}{5}$, $a_2 = \frac{2}{25}$, $a_3= -\frac{1}{2}$. After that, we
obtain
\begin{eqnarray}\label{Mel16}
&{}& \stackrel{2}{\lambda}{} ^{\alpha\gamma} \lambda_{\gamma} - \frac{3}{5} Q_1
\lambda^{\alpha\gamma}\lambda_{\gamma} + \left[ \frac{2}{25}Q_1^2 -\frac{1}{2}  (Q_2 -
\frac{7}{25} Q_1^2)
\right] \lambda^{\alpha} = \\
&{}& = \frac{c^3}{m_0^5} \left[
\begin{pmatrix}
 X_3  \\
  {}  \\
  0_{i}
\end{pmatrix}
+ \frac{1}{c} \begin{pmatrix}
  0  \\
    {} \\
 V_2^i
\end{pmatrix} + 0 \left( \frac{1}{c} \right) \right] \, . \nonumber
\end{eqnarray}
This 4-vector can replace  $\stackrel{2}{\lambda}{} ^{\alpha\gamma} \lambda_{\gamma}$ in
(\ref{11}); so we find that it, together with the factor $h_2$, contributes to
$\phi'^{k}$ the  term
$H_{2}(-2\lambda^{2}_{kh}\lambda_{hll}+\frac{6}{5}\lambda_{ll}\lambda_{ka}\lambda_{all}+4\lambda_{ka}\lambda_{a}\lambda_{ppll}
-\frac{11}{25}\lambda^{2}_{ll}\lambda_{kll}-\lambda_{kll}\lambda_{a}\lambda_{all}+\lambda_{k}\lambda_{all}\lambda_{all}+
+ (tr \lambda^{2}_{ab})\lambda_{kll}-\frac{12}{5}\lambda_{ppll}\lambda_{ll}\lambda_{k})$
and to $h'$ the term $H_{2}X_3$, with $H_2$
limit of $\frac{c^3}{m_0^8}h_2$ for $c$ going to infinity. \\
Finally, we search the numbers $a_4$, $a_5$, $a_6$, $a_7$, $a_8$, $a_9$, such that \\
$\frac{m_0^7}{c^5} \left\{ \right. \stackrel{3}{\lambda}{} ^{\alpha\gamma}
\lambda_{\gamma} + a_4 Q_1 \stackrel{2}{\lambda}{} ^{\alpha\gamma} \lambda_{\gamma}+
\left[ a_5 Q_1^2+ a_6 (Q_2 - \frac{7}{25} Q_1^2) \right]
\lambda^{\alpha\gamma}\lambda_{\gamma} $ \\
$+ \left[ a_7Q_3 + a_8 Q_1^3+ a_9 Q_1  (Q_2 - \frac{7}{25} Q_1^2) \right]
\lambda^{\alpha} \left. \right\} $, has zero limit for $c$ going to infinity. We find
$a_4= - \frac{4}{5}$, $a_5 = \frac{1}{5}$, $a_6= -\frac{1}{2}$, $a_7= - \frac{1}{3}$,
$a_8 = \frac{1}{75}$, $a_9= \frac{2}{5}$. After that we have the result
\begin{eqnarray}\label{Mel17}
&{}& \stackrel{3}{\lambda}{} ^{\alpha\gamma} \lambda_{\gamma} - \frac{4}{5} Q_1
\stackrel{2}{\lambda}{} ^{\alpha\gamma} \lambda_{\gamma}+ \left[ \frac{1}{5} Q_1^2
-\frac{1}{2} (Q_2 -
\frac{7}{25} Q_1^2) \right] \lambda^{\alpha\gamma}\lambda_{\gamma}+ \\
&{}&  \left[ - \frac{1}{3} Q_3 + \frac{1}{75} Q_1^3+ \frac{2}{5} Q_1  (Q_2 -
\frac{7}{25} Q_1^2) \right] \lambda^{\alpha} =  \nonumber \\
&{}& = \frac{c^3}{m_0^7} \left[
\begin{pmatrix}
 X_4  \\
  {}  \\
  0_{i}
\end{pmatrix}
+ \frac{1}{c} \begin{pmatrix}
  0  \\
    {} \\
V_3^i
\end{pmatrix} + 0 \left( \frac{1}{c} \right) \right] \, . \nonumber
\end{eqnarray}
It may replace $\stackrel{3}{\lambda}{} ^{\alpha\gamma} \lambda_{\gamma}$ in (\ref{11}),
so that it, together with the factor $h_3$, contributes to $\phi'^{k}$ the term $H_{3}
\left[ 2\lambda_{ppll} \left( -2\lambda^{2}_{kh}\lambda_{h} - tr \lambda^{2}_{ab}
\lambda_{k}-\frac{8}{5}\lambda_{ll}\lambda_{ka}\lambda_{a}+
\frac{17}{25}\lambda^{2}_{ll}\lambda_{k} \right) + (\lambda_{kh}\lambda_{h})(\lambda_{all}\lambda_{all})+\right. $ \\
$ - \frac{4}{5}\lambda_{ll}
(\lambda_{all}\lambda_{all})\lambda_{k}-\frac{17}{25}\lambda^{2}_{ll}\lambda_{ka}\lambda_{all}
-(\lambda_{all}\lambda_{all})\lambda_{kb}\lambda_{bll}+(tr
\lambda^{2}_{ab})\lambda_{kc}\lambda_{cll}+ \frac{4}{5}\lambda_{ll}
(\lambda_{a}\lambda_{all})\lambda_{kll}+\frac{8}{5}\lambda_{ll}\lambda^{2}_{kh}\lambda_{hll}+$ \\
$+ \frac{74}{375}\lambda^{3}_{ll}\lambda_{kll}  - \frac{4}{5} \lambda_{ll}(tr
\lambda^{2}_{ab}) \lambda_{kll}+ (\lambda_{ab} \lambda_{all} \lambda_{bll})\lambda_{k} -
\left.  (\lambda_{ab} \lambda_{a} \lambda_{bll})\lambda_{kll} + \frac{2}{3} (tr
\lambda^{3}_{ab}) \lambda_{kll}-2\lambda^{3}_{kh}\lambda_{hll} \right] $ and to $h'$ the
term $H_{3}X_4$, with $H_3$ limit of $\frac{c^3}{m_0^8}h_3$ for $c$ going to infinity.
\subsection{The  scalars P$_0$ - P$_3$ .}
From eq. $(\ref{Mel10})_2$ we find $P_0 = \frac{c^6}{m_0^2} \left( 64 \lambda_{ppll}^2 +
0(\frac{1}{c}) \right)$ so that the limit of  $\frac{m_0^2}{c^6} P_0$ is a function of
$X_1$. In order to obtain a less restrictive result, we may substitute $P_0$ with $P_0+
\frac{4}{25}Q_1^2m_0^2c^2$, because this will eliminate the term  $\frac{c^6}{m_0^2} 64
\lambda_{ppll}^2$ di $P_0$; in fact we obtain
\begin{eqnarray*}
P_0+ \frac{4}{25}Q_1^2m_0^2c^2 = \frac{c^4}{m_0^2} \left( -2 X_2 + 0(\frac{1}{c}) \right)
\, .
\end{eqnarray*}
In this way we have a better result, even if it is still not enough; to this end let us
substitute  $P_0+ \frac{4}{25}Q_1^2m_0^2c^2 $ with $P_0+ \frac{4}{25}Q_1^2m_0^2c^2 +2
(Q_2 - \frac{7}{25} Q_1^2)m_0^2c^2 $ and the result is satisfactory because is equal to
\begin{eqnarray*}
\frac{c^2}{m_0^2} \left( - \frac{2}{5} \lambda^{2}_{ll} +  16 \lambda_{ppll} \Lambda - 4
\lambda_{a} \lambda_{all} + 2 tr \lambda^{2}_{ab}  + 0(\frac{1}{c}) \right) \, .
\end{eqnarray*}
In this way we have found the new scalar  $X_5$. \\
Let us now consider the scalar  $P_1$ in $(\ref{13})_6$. It is obvious, for eq.
(\ref{Mel15}) that it is better to replace it with $P_1- \frac{2}{5} Q_1P_0$; but this
has limit a scalar which is a function of $X_1$, $X_2$, $X_3$. Briefly, let us look for
the numbers  $b_1$ and $b_2$ such that
\begin{eqnarray*}
\frac{m_0^4}{c^4} \left\{ P_1- \frac{2}{5} Q_1P_0 +b_1 Q_1 \left( Q_2 - \frac{7}{25}
Q_1^2 \right) m_0^2c^2 +b_2 \left[ Q_3 - \frac{9}{10} Q_1(Q_2 - \frac{7}{25} Q_1^2) -
\frac{11}{125} Q_1^3 \right] m_0^2c^2 \right\}
\end{eqnarray*}
has zero limit. \\
We find $b_1= \frac{2}{5}$, $b_2=\frac{4}{3}$. After that we have
\begin{eqnarray*}
P_1 &-& \frac{2}{5} Q_1P_0 +\frac{2}{5} Q_1 \left( Q_2 - \frac{7}{25} Q_1^2 \right)
m_0^2c^2 +\frac{4}{3} \left[ Q_3 - \frac{9}{10} Q_1(Q_2 - \frac{7}{25} Q_1^2) -
\frac{11}{125} Q_1^3 \right]
m_0^2c^2= \\
&{}& \\
&=& \frac{c^2}{m_0^4} \left( X_6  + 0(\frac{1}{c}) \right) \, ,
\end{eqnarray*}
from which the new scalar $X_6$. \\
Let s consider now the scalar $P_2$ in $(\ref{13})_7$. It is obvious, for eq.
(\ref{Mel16}) that it is better replace it with $P_2- \frac{3}{5} Q_1P_1+ \left[
\frac{2}{25}Q_1^2 -\frac{1}{2}  \left( Q_2 - \frac{7}{25} Q_1^2 \right) \right]P_0$; but
also this is not enough; then we search the numbers   $b_3$ and $b_4$ such that
\begin{eqnarray*}
\frac{m_0^6}{c^4} \left\{ P_2- \frac{3}{5} Q_1P_1+ \left[ \frac{2}{25}Q_1^2 -\frac{1}{2}
\left( Q_2 - \frac{7}{25} Q_1^2 \right) \right]P_0 +  \right. \\
+ b_3Q_1 \left. \left[ Q_3 - \frac{9}{10}
Q_1 \left( Q_2 - \frac{7}{25} Q_1^2 \right) - \frac{11}{125} Q_1^3 \right] m_0^2c^2  \right\} + \\
+b_4 \frac{m_0^6}{c^4} \left\{  Q_4 - \frac{16}{15} Q_1 \left[ Q_3 - \frac{9}{10} Q_1
\left( Q_2 -
\frac{7}{25} Q_1^2 \right) - \frac{11}{125} Q_1^3 \right] + \right. \\
- \frac{1}{2}\left.  \left( Q_2 - \frac{7}{25} Q_1^2 \right)^2   - \frac{14}{25} Q_1^2
\left( Q_2 - \frac{7}{25} Q_1^2 \right) - \frac{19}{625} Q_1^4   \right\} m_0^2c^2
\end{eqnarray*}
has zero limit. \\
We find $b_3= \frac{4}{15}$, $b_4=1$. After that it follows
\begin{eqnarray*}
 P_2- \frac{3}{5} Q_1P_1+ \left[ \frac{2}{25}Q_1^2 -\frac{1}{2}
\left( Q_2 - \frac{7}{25} Q_1^2 \right) \right]P_0 +   \\
+ \frac{4}{15}Q_1  \left[ Q_3 - \frac{9}{10}
Q_1 \left( Q_2 - \frac{7}{25} Q_1^2 \right) - \frac{11}{125} Q_1^3 \right] m_0^2c^2   + \\
+ \left\{  Q_4 - \frac{16}{15} Q_1 \left[ Q_3 - \frac{9}{10} Q_1 \left( Q_2 -
\frac{7}{25} Q_1^2 \right) - \frac{11}{125} Q_1^3 \right] \right. +  \\
- \frac{1}{2} \left. \left( Q_2 - \frac{7}{25} Q_1^2 \right)^2   - \frac{14}{25} Q_1^2
\left(
Q_2 - \frac{7}{25} Q_1^2 \right) - \frac{19}{625} Q_1^4   \right\} m_0^2c^2= \\
{} \\
= \frac{c^2}{m_0^6} \left( X_7  + 0(\frac{1}{c}) \right) \, ,
\end{eqnarray*}
from which the new scalar $X_7$. \\
Finally, let us consider the  scalar $P_3$ in $(\ref{13})_8$. It is obvious, for eq.
(\ref{Mel17}) that it is better replace it with $P_3 - \frac{4}{5} Q_1 P_2+ \left[
\frac{1}{5} Q_1^2 -\frac{1}{2} (Q_2 - \frac{7}{25} Q_1^2) \right] P_1 + \left[ -
\frac{1}{3} Q_3 + \frac{1}{75} Q_1^3+ \frac{2}{5} Q_1  (Q_2 - \frac{7}{25} Q_1^2) \right]
P_0$; but this also is not enough; then we look for a number  $b_5$ such that
\begin{eqnarray*}
\frac{m_0^8}{c^6} \left\{ P_3 - \frac{4}{5} Q_1 P_2+ \left[ \frac{1}{5} Q_1^2
-\frac{1}{2} (Q_2 - \frac{7}{25} Q_1^2) \right] P_1 + \left[ - \frac{1}{3} Q_3 +
\frac{1}{75} Q_1^3+ \frac{2}{5} Q_1  (Q_2 - \frac{7}{25} Q_1^2) \right] P_0
  \right\} + \\
+b_5 Q_1 \frac{m_0^8}{c^6} \left\{  Q_4 - \frac{16}{15} Q_1 \left[ Q_3 - \frac{9}{10} Q_1
\left( Q_2 -
\frac{7}{25} Q_1^2 \right) - \frac{11}{125} Q_1^3 \right] + \right. \\
- \frac{1}{2}\left.  \left( Q_2 - \frac{7}{25} Q_1^2 \right)^2   - \frac{14}{25} Q_1^2
\left( Q_2 - \frac{7}{25} Q_1^2 \right) - \frac{19}{625} Q_1^4   \right\} m_0^2c^2
\end{eqnarray*}
has zero limit. \\
We find $b_5=\frac{1}{5}$. Then we can evaluate
\begin{eqnarray*}
 P_3 - \frac{4}{5} Q_1 P_2+ \left[ \frac{1}{5} Q_1^2
-\frac{1}{2} (Q_2 - \frac{7}{25} Q_1^2) \right] P_1 + \left[ - \frac{1}{3} Q_3 +
\frac{1}{75} Q_1^3+ \frac{2}{5} Q_1  (Q_2 - \frac{7}{25} Q_1^2) \right] P_0
   + \\
+\frac{1}{5} Q_1  \left\{  Q_4 - \frac{16}{15} Q_1 \left[ Q_3 - \frac{9}{10} Q_1 \left(
Q_2 -
\frac{7}{25} Q_1^2 \right) - \frac{11}{125} Q_1^3 \right] + \right. \\
- \frac{1}{2}\left.  \left( Q_2 - \frac{7}{25} Q_1^2 \right)^2   - \frac{14}{25} Q_1^2
\left( Q_2 - \frac{7}{25} Q_1^2 \right) - \frac{19}{625} Q_1^4   \right\} m_0^2c^2
= \\
{} \\
= \frac{c^2}{m_0^8} \left( X_8  + 0(\frac{1}{c}) \right) \, ,
\end{eqnarray*}
from which the last new scalar $X_8$. \\
${} $ \\
\textbf{NOTE:} We have obtained the decomposition (\ref{Mel10}) by considering a modified
procedure of that introduced in \cite{10}, \cite{11} for ideal gases; instead of this, if
we use exactly the procedure of  \cite{10}, \cite{11}, we have to substitute
(\ref{Mel10}) with
\begin{eqnarray}\label{Mel18}
{\lambda^\beta}_\gamma&=& \frac{c^2}{m_0^2} \left[
\begin{pmatrix}
  -8 \lambda_{ppll} & 0_{j} \\
  {} & {} \\
  0_{i} & -4 \lambda_{ppll} \delta_{ij}
\end{pmatrix} + \frac{1}{c} \begin{pmatrix}
  0 & - \lambda_{jll} \\
    {} & {} \\
  \lambda_{ill} &  0_{ij}
\end{pmatrix} +  \right. \\
&{}& \nonumber \\
&+& \frac{1}{c^2} \left. \begin{pmatrix}
  0 & 0_{j} \\
    {} & {} \\
  0_{i} & \lambda_{ij}
\end{pmatrix}+ \frac{1}{c^3}  \begin{pmatrix}
0 & - \frac{1}{2}\lambda_{j} \\
    {} & {} \\
  \frac{1}{2} \lambda_{i} & 0_{ij}
\end{pmatrix}
+ \frac{1}{c^4}  \begin{pmatrix}
  - \lambda & 0_{j} \\
    {} & {} \\
  0_{i} & 0_{ij}
\end{pmatrix} \right]  \, , \nonumber \\
&{}& \nonumber \\
{\lambda^\beta}&=& \frac{c^3}{m_0} \left[
\begin{pmatrix}
  8 \lambda_{ppll}  \\
  {}  \\
  0_{i}
\end{pmatrix} + \frac{1}{c} \begin{pmatrix}
  0  \\
    {} \\
  -2 \lambda_{ill}
\end{pmatrix}  \right]  \, . \nonumber
\end{eqnarray}
Another  procedure is present in literature ( \cite{12}, \cite{13}) also for ideal gases.
If we want to follow it, then we have to substitute (\ref{Mel10}) with
\begin{eqnarray}\label{Mel19}
{\lambda^\beta}_\gamma&=& \frac{c^2}{m_0^2} \left[
\begin{pmatrix}
  -3 \lambda_{ppll} & 0_{j} \\
  {} & {} \\
  0_{i} &  \lambda_{ppll} \delta_{ij}
\end{pmatrix} + \frac{1}{c} \begin{pmatrix}
  0 & - \lambda_{jll} \\
    {} & {} \\
  \lambda_{ill} &  0_{ij}
\end{pmatrix} +
+ \frac{1}{c^2}  \begin{pmatrix}
  0 & 0_{j} \\
    {} & {} \\
  0_{i} & \lambda_{<ij>}
\end{pmatrix}\right]  \, ,  \\
&{}& \nonumber \\
{\lambda^\beta}&=& \frac{c^3}{m_0} \left[
\begin{pmatrix}
  8 \lambda_{ppll}  \\
  {}  \\
  0_{i}
\end{pmatrix} + \frac{1}{c} \begin{pmatrix}
  0  \\
    {} \\
  -2 \lambda_{ill}
\end{pmatrix} + \frac{1}{c^2}  \begin{pmatrix}
  - \frac{2}{3} \lambda_{ll}  \\
    {}  \\
  0_{i}
\end{pmatrix} + \frac{1}{c^3}  \begin{pmatrix}
 0 \\
    {}\\
  \lambda_{i}
\end{pmatrix} \right]  \, , \nonumber \\
\xi &=& c^4 \left( 5 \lambda_{ppll} - \frac{2}{3} \frac{1}{c^2} \lambda_{ll} +
\frac{\lambda}{c^4} \right) \, . \nonumber
\end{eqnarray}
We have performed calculations also with (\ref{Mel18}) and with (\ref{Mel19}) instead of
(\ref{Mel10}), but we don' t report them for the sake of brevity. The interesting result
is that  the pertinent  polynomials in $1/c$ are different between them and from those
here obtained, but the limits for $c$ going to infinity are the same with all 3
approaches!

\end{document}